\documentclass[letterpaper,journal]{IEEEtran}
\usepackage{amsmath,amssymb,amsfonts,amsthm}
\usepackage{algorithmic}
\usepackage{algorithm}
\usepackage{array}
\usepackage[caption=false,font=footnotesize,labelfont=rm,textfont=rm]{subfig}
\usepackage{textcomp}
\usepackage{stfloats}
\usepackage{url}
\usepackage{verbatim}
\usepackage{graphicx}
\usepackage{cite}
\usepackage{xcolor}
\newlength{\simfigurewidth}
\definecolor{red}{RGB}{255,0,0}
\usepackage{cuted}
\usepackage[colorlinks=true,
linkcolor=blue,
citecolor=blue,
urlcolor=blue]{hyperref}
\newtheorem{lemma}{Lemma}

\newtheorem{proposition}{Proposition}
\newtheorem{assumption}{Assumption}

\begin{document}

\title{Robust Beamforming and Antenna Position Optimization for MA-Assisted ISAC with Imperfectly Positioned MAs}

%\author{IEEE Publication Technology,~\IEEEmembership{Staff,~IEEE,}
%        % <-this % stops a space
%\thanks{This paper was produced by the IEEE Publication Technology Group. They are in Piscataway, NJ.}% <-this % stops a space
%\thanks{Manuscript received April 19, 2021; revised August 16, 2021.}}
%\author{
%	\IEEEauthorblockN{Zeyuan Zhang{$^\dagger$},~Yue Xiu{$^\dagger$},~Boyang Jia{$^*$},~Ning Wei{$^\dagger$}}\\
%	\IEEEauthorblockA{{$^\dagger$}{National Key Laboratory of Wireless Communications,} \\
%		{University of Electronic Science and Technology of China, Chengdu, China }\\
%		{{$^*$}National Key Laboratory of Radar Signal Processing, Xidian University, Xi'an, China.
%		}\\
%		Email: wn@uestc.edu.cn
%	}
%}
\author{
	\IEEEauthorblockN{Zeyuan Zhang,~Ning Wei,~Ahmad Bazzi,~Boyang Jia,~Huimin Tang,~Huai Wang and Yue Xiu}\\
	\thanks{
		Z. Zhang, N. Wei and Y. Xiu are with the National Key Laboratory of Wireless Communications, University of Electronic Science and Technology of China, Chengdu 611731, China (e-mail: zzycu@std.uestc.edu.cn; wn@uestc.edu.cn; xiuyue12345678@163.com).}
	\thanks{Ahmad Bazzi is with the Engineering Division, New York University (NYU) Abu Dhabi, Abu Dhabi, United Arab Emirates (e-mail:ahmad.bazzi@nyu.edu).}
		\thanks{B. Jia is with the National Key Laboratory of Radar Signal Processing, Xidian University, Xi'an, 710071, China.
			boyangjia@stu.xidian.edu.cn}
		\thanks{Huimin Tang is with the School of Transportation and Logistics, Southwest Jiaotong University (e-mail: tanghuimin@my.swjtu.edu.cn)}
		\thanks{Huai Wang is with the Space Star Technology Co., Ltd., Beijing 100095, China (e-mail: wanghuai4210474@163.com)}	
}
% The paper headers
%\markboth{Journal of \LaTeX\ Class Files,~Vol.~14, No.~8, August~2021}%
%{Shell \MakeLowercase{\textit{et al.}}: A Sample Article Using IEEEtran.cls for IEEE Journals}
%
%\IEEEpubid{0000--0000/00\$00.00~\copyright~2021 IEEE}
% Remember, if you use this you must call \IEEEpubidadjcol in the second
% column for its text to clear the IEEEpubid mark.

\maketitle

\begin{abstract}
This paper investigates robust beamforming and antenna position design for movable antenna (MA)-enabled integrated sensing and communication (ISAC) systems under antenna position errors. In particular, deviations between the actual and nominal antenna positions in practical MA arrays are inevitable due to limited mechanical accuracy. Therefore, the conventional ISAC design based on nominal positions may suffer from substantial performance degradation.
To address this issue, we consider a monostatic downlink  MA-ISAC system in which each transmission frame consists of a low-power pilot and an ISAC payload. The target angle, reflection coefficient, and antenna position-error vector are jointly estimated based on these two observations, thereby enabling sensing with imperfectly known antenna positions. 
%Since the residual antenna position uncertainty simultaneously affects the sensing accuracy and communication reliability, 
We derive the resulting angle Cramér–Rao bound (CRB) and formulate a problem that minimizes its worst-case value subject to each user's signal-to-interference-plus-noise ratio (SINR) requirement under all possible position errors. By exploiting a common phase-response decomposition, we obtain tractable reformulations and develop an alternating optimization (AO) algorithm for the beamforming vectors and nominal antenna positions. 
Numerical results demonstrate that, compared with conventional nominal-position designs, the proposed robust scheme achieves improved communication reliability while maintaining competitive sensing accuracy under practical antenna position uncertainty

\end{abstract}

\section{Introduction}
The sixth-generation (6G) wireless networks are envisioned to support a wide range of sensing-enabled applications, such as autonomous transportation, unmanned aerial vehicle (UAV) coordination, and intelligent robotic systems. These emerging applications impose increasingly stringent requirements on both communication quality and environmental sensing capability. Motivated by this demand, integrated sensing and communication (ISAC) has recently emerged as a promising technology that enables sensing and communication functionalities to share spectrum, radio-frequency hardware, and signal-processing resources \cite{liu2022isac, ma2026survey}. 
High-performance ISAC systems generally rely on well-designed antenna arrays to provide degrees of freedom for interference suppression and target illumination. However, conventional fixed-position antenna (FPA) arrays are limited by their predetermined array geometry, which restricts their capability to adapt the spatial channel to dynamically varying propagation environments. To overcome this limitation, movable antennas (MAs) have attracted considerable attention by introducing antenna positions as additional design variables. Specifically, an MA array can reconfigure its antenna elements within a prescribed movement region to reshape the wireless channel \cite{zhu2024ma,zhu2025matutorial}. Existing experimental studies have also demonstrated that wavelength-scale antenna movement can lead to significant variations in beampattern \cite{dong2024prototype}. These characteristics make MAs particularly attractive for ISAC systems. 
\subsection{Related Works}
\subsubsection{ISAC}
The concept of ISAC has evolved from the early coexistence of communication and radar systems toward the joint design of dual-functional transceivers.
Recent ISAC studies have examined sensing accuracy, communication performance,
and their interaction under different signaling and deployment settings.  For example, the
work in \cite{ren2024fundamental} characterized the Cram\'er--Rao bound
(CRB)--rate region for
a multi-target ISAC system, with and without prior
target knowledge.  Random information-bearing signals for ISAC were considered in
\cite{lu2024random}, where the ergodic linear minimum mean-square error was
introduced and both data-dependent and data-independent precoders were developed.
For a multi-cell anti-UAV scenario, the authors of
\cite{zhang2025cooperative} jointly optimized the transmit and receive
beamformers to maximize the sensing signal-to-clutter-plus-noise ratio  and
provided both centralized and distributed solvers. 
% In near-field wideband ISAC,
%\cite{zhang2025nearfield} derived angle-and-distance CRBs and employed
%delay--phase precoding to mitigate beam squint effect under a communication rate
%requirement.
Under uncertain target locations, \cite{lan2026robust}
minimized the worst-case localization CRB over a coarse location area for
multiuser multiple-input multiple-output orthogonal frequency-division
multiplexing (MIMO--OFDM) ISAC through semidefinite relaxation.  A clutter-aware
framework in \cite{luo2024clutter} combined communication beams with a
rotating sensing beam and estimated the angle, range, and velocity of dynamic
targets after suppressing static clutter.  For cooperative bistatic ISAC,
\cite{ren2026twotimescale} adapted access point (AP) modes and power allocation
to statistical channel state information (CSI) while updating the beamformers
with instantaneous CSI.  
\subsubsection{MA Position Error}
Finite positioning accuracy is an intrinsic implementation issue for MA
systems: owing to mechanical resolution and calibration mismatch, the realized
antenna position vector (APV) can deviate from its optimized or commanded value.
Such errors directly distort the steering phase and may
therefore impair both communication beamforming and sensing.
The work in \cite{su2025maerrors} characterized the worst-case
near-field beam gains for beam nulling and multibeam formation under bounded MA
position errors using a Taylor approximation.  The
subsequent study jointly designed the antenna weights and positions and
further quantified the sensitivity of the optimized near-field beamforming
solutions to positioning errors \cite{yang2026flexible}.  Finite positioning
precision has also been handled through discrete MA locations in a two-timescale
MA-ISAC design \cite{khalili2026twotimescale} and analyzed in a sparse fluid
antenna system (FAS) \cite{wu2026sparsefas}.  The self-calibration method in
\cite{ye2026selfcalibration} alternated between multiple signal classification
(MUSIC)-based DOA estimation and
closed-form estimation of the unknown MA position errors.  A 300-GHz
measurement platform in \cite{wang2024mameasurement} used a two-dimensional
displacement system with 0.02-mm precision to characterize channels over
$32\times32$ candidate MA ports. 
 Motion-induced gain and phase errors were
incorporated into MA-array DOA estimation in \cite{ye2025gainphase}.  A recent position-calibration method in
\cite{liu2026arraycalibration} jointly
estimated DOAs, angular spreads, and sensor-position perturbations for
partially calibrated arrays through weighted least squares and alternating
optimization.
Classical array-shape calibration
provides the broader identifiability background for jointly estimating source
directions and unknown sensor locations \cite{rockah1987near}.
\subsubsection{MA-Enabled ISAC}
Extensive research has shown that MA can effectively
promote beamforming gain and increasing attention has been
paid to employing MA to facilitate ISAC 
\cite{lyu2025maisac,jiang2025maisac,chen2025maisac,
wang2024fluidisac,tang2026fullduplex,khalili2026twotimescale, zhang2026crosstalk}.  In
\cite{lyu2025maisac}, beamformer and APV were designed using
communication rate and sensing mutual information in a bistatic multiuser
multiple-input single-output (MU-MISO) system. Authors in  \cite{jiang2025maisac} placed MAs at both the transmit and
receive sides of the base station (BS) and maximized the sensing
signal-to-interference-plus-noise ratio (SINR) under
communication requirements.  \cite{chen2025maisac} derived the angle CRB and
developed closed-form and search-based solutions for receive- and transmit-MA
settings.  For FAS-assisted multiuser ISAC, \cite{wang2024fluidisac} used deep
reinforcement learning for joint port selection and precoding under a sensing
constraint.  \cite{tang2026fullduplex} jointly optimized the transmit and
receive FAS positions, beamformers, and uplink power in a full-duplex system. 
Under antenna crosstalk, \cite{zhang2026crosstalk} generalized the FPA
crosstalk model to movable arrays and jointly optimized the beamformers and MA
positions to minimize the sensing CRB using deep reinforcement learning.

% In a non-orthogonal multiple access (NOMA)-based architecture,
%\cite{lyu2026nomasensing} jointly designed the power
%allocation, beamformers, and two-dimensional MA positions to increase
%multi-target illumination while meeting user SINR requirements.  
%Secure
%intelligent reflecting surface (IRS)-assisted ISAC was considered in
%\cite{cao2026secureirs}, which optimized the communication beamformer, sensing
%covariance, IRS phase shifts, and MA
%positions. 
% For low-altitude networks, \cite{li2026uavmaisac} jointly adjusted
%UAV trajectories, user association, MA positions, and beamforming using
%clustering and soft actor--critic learning.

\subsection{Motivation and Contributions}
In practical MA arrays, finite mechanical accuracy causes the realized APV to
differ from the commanded geometry.  Existing studies have quantified
near-field beamforming losses under position errors
\cite{su2025maerrors,yang2026flexible} and modeled finite precision through
discrete MA-ISAC locations \cite{khalili2026twotimescale}.  Sensing-oriented
work has also jointly estimated target directions and MA position errors
\cite{ye2026selfcalibration}.  These studies establish the importance of
position uncertainty, but leave unresolved how communication and sensing can
be jointly protected when the actual geometry is unknown.
This raises the central question: how can an MA-ISAC system achieve robust
multiuser communication and reliable target-angle sensing when a common
position error simultaneously distorts the user channels and target response?
To address the MA location uncertainty problem for MA-ISAC systems, it is important to develop sensing models that can account for the unknown geometry, instead of evaluating performance at the specified positions. 
Pilot-assisted calibration provides a possible route.  Pilot sources have
been used to identify array-manifold errors involving sensor locations, gains,
phases, and mutual coupling \cite{stavropoulos2000pilotcalibration}.
Near-field observation receivers have also exploited transmitter-side
feedback signals to identify propagation responses for array predistortion
\cite{benayed2022nftor}.  These results suggest that a known low-power
excitation observed through local near-field coupling can carry useful
information about the realized array.
Motivated by this idea, we use low-power orthogonal pilots and direct
transmit-to-sensing leakage to obtain a physical observation of the actual APV.
Together with the target echo, this observation enables joint estimation of
the target angle and MA position error.  It then supports an
observation-consistent sensing limit and a robust spatial design that protects
both sensing and communication against the same bounded position uncertainty.

The main contributions of this paper are summarized as follows.
\begin{itemize}
\item We establish a pilot-assisted multiuser MA-ISAC model with bounded
transmit-antenna position errors.   By jointly exploiting the pilot and target-echo observations, we formulate a unified estimation model for the target angle, reflection coefficient, and APV error.  Based on this model, we derive the corresponding CRB of the sensing angle between the target and BS,  which further reveals that the fixed receive aperture remains directly usable, whereas the contribution of the movable transmit aperture is determined by the pilot information.

\item Next, based on the derived CRB, we formulate a worst-case angle-CRB minimization problem that jointly
optimizes the beamforming matrix and nominal APV.   The resulting design guarantees the SINR requirement of each user while satisfying the transmit-power, antenna-movement, and minimum-spacing constraints for every admissible realization of the antenna position error.

\item Furthermore, to solve the resulting highly non-convex semi-infinite problem, we develop an analytical robust optimization framework. Specifically, a common phase-response decomposition is exploited to derive the tractable finite-dimensional form of  worst-case SINR and sensing requirement.  Based on these transformations, an alternating optimization (AO) algorithm is developed to successively optimize the beamforming vectors and nominal APV while preserving feasibility and monotonically improving the guaranteed sensing performance.

\item Finally, numerical results demonstrate the effectiveness of the proposed robust MA-ISAC design. Compared with conventional nominal-position design, the proposed scheme effectively mitigates the performance degradation caused by antenna position uncertainty and achieves lower angle CRBs while maintaining robust multiuser communication performance.
\end{itemize}

\subsection{Organization and Notation}
The remainder of this paper is organized as follows.
Section~\ref{sec:system_model} presents the considered MA-ISAC system model
and formulates the robust design problem. Section~\ref{sec:proposed_algorithm}
derives the robust reformulations and develops the AO algorithm.
Section~\ref{sec:numerical_results} reports numerical results, and
Section~\ref{sec:conclusion} concludes the paper.

Throughout the paper, scalars, vectors, and matrices are denoted by \(x\), \(\mathbf x\), and
\(\mathbf X\), respectively. Superscripts \((\cdot)^*\),
\((\cdot)^{\mathrm T}\), \((\cdot)^{\mathrm H}\), and
\((\cdot)^\dagger\) denote complex conjugation, transpose, Hermitian
transpose, and the Moore--Penrose pseudoinverse, respectively. The norm
\(\|\cdot\|_2\) is Euclidean for vectors and spectral for matrices, 
\(\|\cdot\|_{\rm F}\) is the Frobenius norm. \(\mathbf I_n\),
\(\mathbf1_n\), and \(\mathbf0\) denote the identity matrix, all-one
vector, and zero vector or matrix of compatible dimensions, respectively.
Moreover, \(\mathbf e_n\) denotes the \(n\)-th canonical vector of compatible
dimension.

\section{System Model}
\label{sec:system_model}

\begin{figure}[t]
\centering
\subfloat[]{%
\includegraphics[width=\columnwidth]{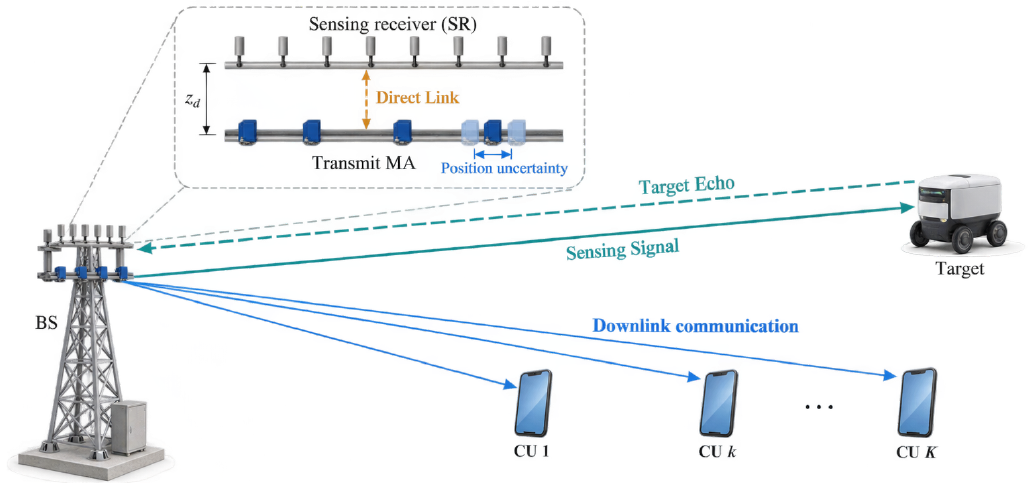}%
\label{fig:isac_system_setting}}\\
\subfloat[]{%
\includegraphics[width=\columnwidth]{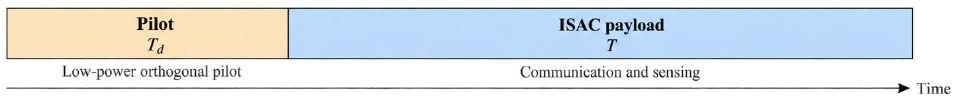}%
\label{fig:isac_frame_structure}}
\caption{The considered MA-ISAC system: (a) system setting and (b) frame structure.}
\label{fig:isac_system_protocol}
\end{figure}

We consider a downlink millimeter-wave system, as illustrated in
Fig.~\ref{fig:isac_system_protocol}(a), in which a
monostatic dual-functional BS simultaneously serves multiple communication
users (CUs) and senses a point target located at angle \(\theta_{\rm t}\).  The BS is equipped with a linear array
of \(N\) transmit MAs, while its co-located sensing receiver (SR) employs \(M\) FPAs
arranged as a uniform linear array (ULA) with \(\frac{\lambda}{2}\) inter-element spacing\footnote{To reveal the fundamental impact of MA position errors on both
		communication and sensing, we employ transmit-side MAs, whose APV jointly
		determines the user channels and target response, while retaining a fixed
		sensing array.}.  The transmit MA array and the receive ULA are parallel and
separated by a perpendicular distance \(z_{\rm d}\).  Besides receiving
the target echo, the SR can also observe the near-field direct link from
the transmit MAs, whose spatial response depends on the MA
geometry.  The CU set is \(\mathcal K=\{1,\ldots,K\}\) with \(K\ge2\), and
each CU is equipped with a single antenna.
Let \(\mathbf p=[p_1,\ldots,p_N]^{\mathrm{T}}\in\mathbb R^{N}\) denote the nominal APV designed by the BS, where \(p_n\) is the intended position of the \(n\)-th MA. Due to practical positioning inaccuracy caused by mechanical  errors, the actual APV deviates from its nominal position, which is modeled as
\begin{equation}
	\tilde{\mathbf p}
	=
	\mathbf p+\Delta\mathbf p,
	\label{eq:actual_apv}
\end{equation}
where \(\Delta\mathbf p=[\Delta p_1,\ldots,\Delta p_N]^{\mathrm{T}}\in\mathbb R^{N}\) denotes the antenna position error vector. We assume each error is bounded by \(\varepsilon\), which gives the position uncertainty set
\(\mathcal E_{\rm p}(\varepsilon)\triangleq
\{\Delta\mathbf p:|\Delta p_n|\le\varepsilon,\ \forall n\}\). More importantly, 
the SR does not have access to the realization of
\(\Delta\mathbf p\) and hence does not know the actual APV
\(\tilde{\mathbf p}\).
In this paper, we focus on the 
small-error regime specified by the following assumption.
\begin{assumption}
\label{ass:small_position_error}
The antenna position error bound satisfies
\(0\le\varepsilon<\lambda/4\).
\end{assumption}
The assumption \ref{ass:small_position_error} indicates that the width of the uncertainty interval is smaller
than half a wavelength \(\frac{\lambda}{2}\) and is consistent with the small position
errors in practical scenarios \cite{su2025maerrors,yang2026flexible}. 
%For any
%angle of departure (AoD) \(\theta\), the phase perturbation of the \(n\)-th MA
%satisfies
%\(\left|(2\pi/\lambda)\cos\theta\,\Delta p_n\right|
%\le(2\pi/\lambda)\varepsilon<\pi/2\)
%and each individual phase factor lies on a symmetric arc shorter than a
%semicircle.
To guarantee that the actual APV remains inside the moving region and satisfies the minimum-spacing requirement for all admissible position errors, the nominal APV is restricted to the following movable region:
\begin{equation}
	\begin{aligned}
	\mathcal P_{\rm r}
	\triangleq
	\{\mathbf p:\;&
	\varepsilon\le p_1<\cdots<p_N
	\le D_{\max}-\varepsilon,
	\\
	&
	p_{n+1}-p_n\ge d_{\min}+2\varepsilon
	\}.
	\end{aligned}
	\label{eq:robust_apv_set}
\end{equation}
Here, \(D_{\max}\) is the length of the MA moving region, \(d_{\min}\) is
the minimum spacing between adjacent MAs.
%The set \(\mathcal P_{\rm r}\) is nonempty if
%\begin{equation}
%	D_{\max}\ge (N-1)d_{\min}+2N\varepsilon.
%	\label{eq:Pr_nonempty_condition}
%\end{equation}
\subsection{ISAC Protocol}
\label{subsec:isac_protocol}

 Before each downlink transmission frame, the beamformer and the nominal APV are first configured.  Once the antenna
positioning process is completed, the MAs remain fixed at an actual APV \(\tilde{\mathbf p}\) deviated from the nominal design \({\mathbf p}\)
 throughout the frame. 
% In the considered ISAC system, the same actual APV determines both the user channels and the
%transmit-side target response.
 To better understand the sensing process, let
\(\kappa\triangleq2\pi/\lambda\) denote the carrier wavenumber. In the
sensing echo, the transmit-side coefficient associated with MA \(n\)
contains the phase factor
\(e^{-j\kappa(p_n+\Delta p_n)\cos\theta_{\rm t}}\).
Since the position error $\Delta p_n$ is unknown,
an angular variation of the target can be locally compensated by corresponding
variations in the MA position errors, which leads to unresolvable ambiguity in target sensing. To deal with this coupling, during each frame, the BS uses a low-power orthogonal pilot before the ISAC payload. As illustrated in
Fig.~\ref{fig:isac_system_protocol}(b), specifically, the MAs send 
\(\mathbf X_{\rm d}\in\mathbb C^{N\times T_{\rm d}}\) over
\(T_{\rm d}\ge N\) snapshots with 
\begin{equation}
	\mathbf X_{\rm d}\mathbf X_{\rm d}^{\mathrm H}
	=
	T_{\rm d}P_{\rm d}\mathbf I_N,
	\label{eq:pilot_covariance}
\end{equation}
where \(P_{\rm d}\) is the pilot power per MA. The SR utilizes the received pilot from the direct transmit-to-sensing link as a transmit geometry observation. Owing to its short one-way propagation, it remains
observable even at low pilot power. 
%Since the
%sensing-array geometry and the array separation \(z_{\rm d}\) are known, each
%transmit MA produces a position-dependent spherical-wave signature across the
%sensing FPA.  Orthogonal pilot excitation allows these signatures to be
%separated across the transmit antennas.

At the payload stage, the BS uses the data
symbols simultaneously for downlink communication and target sensing over the
same time-frequency resource. Let
\(\mathbf s[t]\in\mathbb C^K\) collect the unit-power data symbols at snapshot
\(t\), with \(s_k[t]\) denoting its \(k\)-th entry.  For a payload of
\(T\ge K\) snapshots, define
\(\mathbf S=[\mathbf s[1],\ldots,\mathbf s[T]]\), with
\(\mathbf S\mathbf S^{\mathrm H}=T\mathbf I_K\). Let
\(\mathbf W=[\mathbf w_1,\ldots,\mathbf w_K]\in\mathbb C^{N\times K}\)
denote the beamforming matrix, where
\(\mathbf w_k=[w_{1,k},\ldots,w_{N,k}]^{\rm T}\) is its \(k\)-th column. The beamformed signal transmitted by the BS is then given by
\(\mathbf X=\mathbf W\mathbf S\).
After both observations have been collected by the SR, the target angle, reflection coefficient, and MA position errors are jointly estimated to perform sensing, which will be detailed in the following. 

\subsection{Signal Model}
\label{subsec:signal_model}

During the pilot stage, the SR observes the known pilot through the
near-field direct link from the transmit MAs. In the considered system model, the \(n\)-th transmit MA
and the \(m\)-th receive antenna are located at \([\tilde p_n,0]^{\rm T}\) and
\([(m-1)\lambda/2,z_{\rm d}]^{\rm T}\), respectively.
According to the near-field spherical-wave model, the \((m,n)\)-th coefficient
of the near-field direct-link channel is modeled as
\begin{equation}
	[\mathbf H_{\rm d}(\tilde{\mathbf p})]_{m,n}
	=
	\frac{\lambda}{4\pi r_{m,n}^{\rm d}(\tilde p_n)}
	e^{-j\kappa r_{m,n}^{\rm d}(\tilde p_n)},
	\label{eq:near_field_direct_channel}
\end{equation}
where \(r_{m,n}^{\rm d}(\tilde p_n)=
\sqrt{(\tilde p_n-(m-1)\lambda/2)^2+z_{\rm d}^2}\) is the physical distance between the \(n\)-th transmit MA
and the \(m\)-th  receive antenna. Here, \(\mathbf H_{\rm d}(\tilde{\mathbf p})\in\mathbb C^{M\times N}\)
denotes the direct-link channel matrix, with \(\mathbf h_{\rm d}(\tilde p_n)\in\mathbb C^M\)  being its \(n\)-th column. In the considered system model, since \(P_{\rm d}\) is deliberately kept low, the pilot echo reflected
by the far-field target after two-way propagation
% is well below the sensing-noise level and
is therefore neglected.  Accordingly, the pilot-stage observation is given by
\begin{equation}
	\mathbf Y_{\rm d}
	=
	\mathbf H_{\rm d}(\tilde{\mathbf p})\mathbf X_{\rm d}
	+
\mathbf Z_{\rm d},
\label{eq:pilot_received_matrix}
\end{equation}
where \(\mathbf Y_{\rm d},\mathbf Z_{\rm d}\in
\mathbb C^{M\times T_{\rm d}}\), \(\mathbf Z_{\rm d}\) is the additive white Gaussian noise (AWGN) whose entries are 
independently distributed as
\(\mathcal{CN}(0,\sigma_{\rm s}^2)\), with $\sigma_{\rm s}^2$ being the sensing noise power.

For an arbitrary AoD \(\theta\), define
the MA steering vector as
\(\mathbf a(\theta,\tilde{\mathbf p})=
[e^{j\kappa\tilde p_1\cos\theta},\ldots,
e^{j\kappa\tilde p_N\cos\theta}]^{\mathrm T}\).
The far-field channel from the BS to CU \(k\) contains \(L_k\) paths. The resulting
channel vector from the BS to the \(k\)-th CU is given by
\begin{equation}
	\mathbf h_k(\tilde{\mathbf p})
	=
	\sum_{\ell=1}^{L_k}
	\alpha_{k,\ell}
	\mathbf a(\theta_{k,\ell},\tilde{\mathbf p}),
	\label{eq:comm_channel}
\end{equation}
where \(\alpha_{k,\ell}\) and \(\theta_{k,\ell}\) denote the complex
gain and AoD of the \(\ell\)-th path of the \(k\)-th CU, respectively.  
Accordingly, the received signal at CU \(k\) is
\begin{equation}
	\begin{aligned}
		y_k[t]
		&=
			\mathbf h_k^{\mathrm{H}}(\tilde{\mathbf p})\mathbf w_k s_k[t]
		+
			\sum_{i\ne k}
			\mathbf h_k^{\mathrm{H}}(\tilde{\mathbf p})\mathbf w_i s_i[t]
		+
			n_k[t],
	\end{aligned}
	\label{eq:received_comm_signal_decomposed}
\end{equation}
where \(n_k[t]\sim\mathcal{CN}(0,\sigma_k^2)\) denotes the AWGN at CU \(k\), and \(\sigma_k^2\) is the noise power.

For the sensing link, let the receive steering vector of the sensing ULA be
\(\mathbf b(\theta_{\rm t})=
[1,e^{j\pi\cos\theta_{\rm t}},\ldots,
e^{j\pi(M-1)\cos\theta_{\rm t}}]^{\mathrm T}\). Further define the target spatial
wavenumber and angular-sensitivity coefficient as
\(\kappa_{\rm t}\triangleq\kappa\cos\theta_{\rm t}\) and
\(\beta_{\rm t}\triangleq\kappa\sin\theta_{\rm t}\). 
For notational simplicity, the target transmit steering vectors evaluated at
the actual and nominal APVs are denoted by
\(\mathbf a_{\rm t}(\tilde{\mathbf p})\triangleq
\mathbf a(\theta_{\rm t},\tilde{\mathbf p})\) and
\(\mathbf a_{\rm t}(\mathbf p)\triangleq
\mathbf a(\theta_{\rm t},\mathbf p)\), respectively.  
The SR records the payload echo observation
over the \(T\) snapshots as
\begin{equation}
	\mathbf Y_{\rm s}
	=
	\alpha_{\rm t}\mathbf b(\theta_{\rm t})
	\mathbf a_{\rm t}^{\rm H}(\tilde{\mathbf p})\mathbf X
	+
	\mathbf Z_{\rm s},
\label{eq:sensing_received_matrix}
\end{equation}
where the target radar cross section (RCS) and propagation attenuation are absorbed
into the complex reflection coefficient \(\alpha_{\rm t}\). Here  $\mathbf Z_{\rm d}$ and $\mathbf Z_{\rm s}$ are independent
AWGN matrices with the same entrywise variance $\sigma_{\rm s}^2$.

\subsection{ISAC Performance Metrics}
\label{subsec:isac_performance_metric}

\begin{figure*}[!b]
\noindent\rule{\textwidth}{0.4pt}
\begin{equation*}
\operatorname{CRB}_{\theta}(\mathbf W,\tilde{\mathbf p})
=
\frac{\sigma_{\rm s}^2}{2T|\alpha_{\rm t}|^2}
\left[
\eta_{\rm r}\|\mathbf u\|_2^2
+M
\frac{
\tilde{\mathbf p}^{\rm T}\mathbf J_{\rm e}
(\mathbf J_{\rm e}+\mathbf J_{\rm d})^{-1}
\mathbf J_{\rm d}\tilde{\mathbf p}
}{
\tilde{\mathbf p}^{\rm T}\mathbf J_{\rm e}\tilde{\mathbf p}
}
\frac{\|\mathbf d\|_2^2}{\|\mathbf u\|_2^2}
\right]
^{-1}.
\tag{\ref*{eq:crb_theta}}
\end{equation*}
\end{figure*}

To evaluate the communication performance under the actual APV \(\tilde{\mathbf p}\), we consider the SINR of CU \(k\), which can be expressed as
\begin{equation}
	\gamma_k(\mathbf W,\tilde{\mathbf p})
	=
	\frac{
		\left|
		\mathbf h_k^{\mathrm{H}}(\tilde{\mathbf p})\mathbf w_k
		\right|^2
	}{
		\sum_{i\ne k}
		\left|
		\mathbf h_k^{\mathrm{H}}(\tilde{\mathbf p})\mathbf w_i
		\right|^2
		+
		\sigma_k^2
	}.
	\label{eq:comm_sinr}
\end{equation}
For the sensing performance, based on the observations in
\eqref{eq:pilot_received_matrix} and \eqref{eq:sensing_received_matrix}, the
unknown parameter \((\theta_{\rm t},\alpha_{\rm t},\Delta\mathbf p)\) can be 
jointly estimated at the SR to conduct target sensing. For example, according to the maximum-likelihood (ML) estimation, \((\theta_{\rm t},\alpha_{\rm t},\Delta\mathbf p)\) can be given by
\begin{equation}
\begin{aligned}
(\widehat\theta_{\rm t},\widehat\alpha_{\rm t},
\Delta\widehat{\mathbf p})
&=
\underset{\theta,\alpha,\Delta\mathbf p}{\operatorname{arg\,min}}
\Big\{
\|\mathbf Y_{\rm d}
-\mathbf H_{\rm d}(\mathbf p+\Delta\mathbf p)\mathbf X_{\rm d}\|_{\rm F}^2
\\
&+
\|\mathbf Y_{\rm s}
-\alpha\mathbf b(\theta)
\mathbf a^{\rm H}(\theta,\mathbf p+\Delta\mathbf p)\mathbf X\|_{\rm F}^2
\Big\}.
\label{eq:joint_ml_estimator}
\end{aligned}\end{equation}
%For each MA \(n\), define the derivative of its direct-link channel vector as
%\(\dot{\mathbf h}_{\rm d}(\tilde p_n)\triangleq
%\partial\mathbf h_{\rm d}(\tilde p_n)/\partial\tilde p_n\).
%Since the pilot covariance in \eqref{eq:pilot_covariance} is
%diagonal, the pilot data provide the diagonal position-information
%matrix
%\begin{equation}
%\label{eq:pilot_position_information}
%	\mathbf J_{\rm d}(\tilde{\mathbf p})
%	\triangleq
%	\operatorname{diag}\!\left(
%	g_{\rm d}(\tilde p_1),\ldots,g_{\rm d}(\tilde p_N)
%	\right),
%\end{equation}
%where
%\(g_{\rm d}(\tilde p_n)
%\triangleq
%\frac{2T_{\rm d}P_{\rm d}}{\sigma_{\rm s}^2}
%\|\dot{\mathbf h}_{\rm d}(\tilde p_n)\|_2^2\). 
The resulting estimates are used to update the target parameters for the next-frame design. Based on the joint observation model described above, the target angle–position error coupling in the payload echo can be resolved using the position error information provided by the pilot. To characterize the accuracy limit for estimating $\theta_{\rm t}$, we consider the CRB for \(\theta_{\rm t}\), in the presence of the nuisance parameters $\alpha_{\rm t}$ and
$\Delta\mathbf p$. To express this bound, first define the effective transmit response as
\(\mathbf u\triangleq
\mathbf W^{\rm H}\mathbf a_{\rm t}(\tilde{\mathbf p})
\). 
Its derivative with respect to the target angle is then
\begin{equation}
\mathbf v\triangleq
\frac{\partial\mathbf u}{\partial\theta_{\rm t}}
=
-j\beta_{\rm t}\mathbf W^{\rm H}
\operatorname{diag}\!\bigl(\mathbf a_{\rm t}(\tilde{\mathbf p})\bigr)
\tilde{\mathbf p}.
\end{equation}
Further let \(\mathbf d\in\mathbb C^{\binom K2}\) stack the pairwise responses
\(d_{i,j}\triangleq v_i u_j-v_j u_i, 1\le i<j\le K\),
in lexicographic order. With these definitions, the following proposition gives the exact form of the angle CRB.
%\begin{proposition}
%\label{prop:rb_exact_aperture_decomposition}
%The angle CRB can be derived as \eqref{eq:crb_theta}, as shown in the bottom of this page, where 
%\(\eta_{\rm r}\triangleq
%\pi^2\sin^2\theta_{\rm t}M(M^2-1)/12\). 
%Finally, $\mathbf J_{\rm e}$ is given by
%\begin{equation}
%\mathbf J_{\rm e}
%\triangleq
%\frac{2TM|\alpha_{\rm t}|^2}{\sigma_{\rm s}^2}
%\Re\!\left\{
%\mathbf G^{\rm H}\mathbf P_{\mathbf u}^{\perp}\mathbf G
%\right\},
%\label{eq:echo_position_information}
%\end{equation}
%where \(\mathbf P_{\mathbf u}^{\perp}\triangleq
%\mathbf I_K-\frac{\mathbf u\mathbf u^{\rm H}}{\|\mathbf u\|_2^2}\) and
%\(\mathbf G\triangleq j\kappa_{\rm t}\mathbf W^{\rm H}
%\operatorname{diag}(\mathbf a_{\rm t}(\tilde{\mathbf p}))\).
%\refstepcounter{equation}\label{eq:crb_theta}
%\end{proposition}
%The proof is given in Appendix~\ref{app:crb_derivation}.  The two terms in
%\eqref{eq:crb_theta} are supplied by the fixed receive aperture and the
%transmit aperture supported by the pilot, respectively.
%  The matrix ratio
%multiplying the transmit-aperture term is its exact retained fraction under
%noisy pilot observations.
\begin{proposition}
	\label{prop:rb_exact_aperture_decomposition}
The angle CRB can be derived as \eqref{eq:crb_theta}, as shown at the bottom of this page, where 
\(\eta_{\rm r}\triangleq
\pi^2\sin^2\theta_{\rm t}M(M^2-1)/12\). The matrix $\mathbf J_{\rm d}$ is given by
	\begin{equation}
		\mathbf J_{\rm d}
		\triangleq
		\operatorname{diag}\!\left(
		g_{\rm d}(\tilde p_1),\ldots,g_{\rm d}(\tilde p_N)
		\right),
		\label{eq:pilot_position_information}
	\end{equation}
	where
	\(	g_{\rm d}(\tilde p_n)
	\triangleq
	\frac{2T_{\rm d}P_{\rm d}}{\sigma_{\rm s}^2}
	\left\|
	\frac{\partial\mathbf h_{\rm d}(\tilde p_n)}
	{\partial\tilde p_n}
	\right\|_2^2\).
The matrix $ \mathbf J_{\rm e}$ is given by
	\begin{equation}
		\mathbf J_{\rm e}
		\triangleq
		\frac{2TM|\alpha_{\rm t}|^2}{\sigma_{\rm s}^2}
		\Re\!\left\{
		\mathbf G^{\rm H}\mathbf P_{\mathbf u}^{\perp}\mathbf G
		\right\},
		\label{eq:echo_position_information}
	\end{equation}
	where
	\(\mathbf P_{\mathbf u}^{\perp}\triangleq
	\mathbf I_K-\frac{\mathbf u\mathbf u^{\rm H}}{\|\mathbf u\|_2^2}\)
	and
	\(\mathbf G\triangleq
	j\kappa_{\rm t}\mathbf W^{\rm H}
	\operatorname{diag}\!\bigl(
	\mathbf a_{\rm t}(\tilde{\mathbf p})
	\bigr)\).
Here, \(\mathbf J_{\rm d}\) and \(\mathbf J_{\rm e}\) denote the pilot and payload information matrices for the MA position errors, respectively. 
\refstepcounter{equation}\label{eq:crb_theta}
\end{proposition}
The proof is given in Appendix~\ref{app:crb_derivation}.
The two terms in \eqref{eq:crb_theta} correspond to the fixed
receive aperture and the MA transmit aperture supported
by the pilot, respectively.
\subsection{Problem Formulation}
\label{subsec:problem_formulation}
Since the actual APV $\tilde{\mathbf p}$ is affected by the unknown position error \(\Delta\mathbf p\), it cannot be designed directly. We therefore jointly optimize the beamforming matrix \(\mathbf W\) and
the nominal APV \(\mathbf p\) for a robust design over the uncertainty set
\(\mathcal E_{\rm p}(\varepsilon)\). Specifically, the objective is to minimize the worst-case
angle CRB among all admissible position error values, while guaranteeing
the user SINR requirements and the MA deployment constraints.
Accordingly, the optimization problem is formulated as
\begin{subequations}
\label{prob:worst_case_crb_minimization}
\begin{align}
	\min_{\mathbf W,\mathbf p}\quad
	&
	\max_{\Delta\mathbf p\in\mathcal E_{\rm p}(\varepsilon)}
	\operatorname{CRB}_{\theta}
	(\mathbf W, \mathbf p+\Delta\mathbf p)
	\label{prob:worst_case_crb_obj}
	\\
	{\rm s.t.}\,\,
	&
	\begin{aligned}[t]
	\gamma_k
	\left(
	\mathbf W,
	{\mathbf p+\Delta\mathbf p}
	\right)
	&\ge
	\Gamma_k,\,
	\forall k\in\mathcal K,\,
	\forall\Delta\mathbf p\in\mathcal E_{\rm p},
	\end{aligned}
	\label{prob:sinr_constraint}
	\\
	&
	\operatorname{tr}
	\left(
	\mathbf W\mathbf W^{\mathrm{H}}
	\right)
	\le
	P_{\max},
	\label{prob:power_constraint}
	\\
	&
	\mathbf p
	\in
	\mathcal P_{\rm r},
	\label{prob:nominal_apv_constraint}
\end{align}
\end{subequations}
where $\Gamma_k$ is the SINR requirement of CU $k$, and \(P_{\max}\) is the
maximum transmit power.  Since the pilot
constitute only a small fraction of the frame power, we use only \eqref{prob:power_constraint} to limit the  transmit power. Meanwhile, constraint \eqref{prob:sinr_constraint}
guarantees the SINR requirement of the CUs under all admissible position errors, and \eqref{prob:nominal_apv_constraint} limits the 
movable region of MA. Notably, \eqref{prob:worst_case_crb_minimization} is difficult to solve for two
reasons. First, the beamformer \(\mathbf W\) and nominal APV
\(\mathbf p\) are nonlinearly coupled, making the problem
highly nonconvex. Second, the worst-case objective and robust SINR constraints involve
a continuum of position errors, making the problem semi-infinite. 

\section{Proposed Robust Beamforming Approach}
\label{sec:proposed_algorithm}

In this section, we propose a robust beamforming approach based on tractable
robust reformulations and AO in order to address problem \eqref{prob:worst_case_crb_minimization}. 
We first rewrite the problem into a more tractable form. Let
\(\gamma_{\rm s}\triangleq2T|\alpha_{\rm t}|^2/\sigma_{\rm s}^2\) and define
the angle information as
\(\mathcal I_\theta(\mathbf W,\tilde{\mathbf p})\triangleq
[\gamma_{\rm s}\operatorname{CRB}_{\theta}
(\mathbf W,\tilde{\mathbf p})]^{-1}\).
Introducing an auxiliary variable \(\tau>0\) as a certified lower bound on 
\(\mathcal I_\theta\), problem~\eqref{prob:worst_case_crb_minimization} can be
equivalently reformulated as
\begin{subequations}
\label{prob:worst_case_information_maximization}
\begin{align}
\max_{\mathbf W,\mathbf p,\tau>0}\quad
&\tau
\label{prob:worst_case_information_objective}\\
\mathrm{s.t.}\quad
&\mathcal I_\theta(\mathbf W,\mathbf p+\Delta\mathbf p)\ge\tau,
\,\,\,\,\forall\Delta\mathbf p\in\mathcal E_{\rm p}(\varepsilon),
\label{prob:worst_case_information_constraint}\\
&\eqref{prob:sinr_constraint},
\eqref{prob:power_constraint},
\eqref{prob:nominal_apv_constraint}.
\end{align}
\end{subequations}
For any feasible solution, the constraint \(\eqref{prob:worst_case_information_constraint}\) guarantees \(\mathcal I_\theta\ge\tau\) for every admissible position error. Accordingly, \(1/(\gamma_{\rm s}\tau)\) is a certified upper bound on the worst-case CRB, which is tight at the optimum of \(\eqref{prob:worst_case_information_maximization}\).
In the following, we first establish a phase-response model shared by the two robust
constraints.  The communication SINR and sensing CRB are then handled in two
separate subsections.  

\subsection{Unified Phase-Response Decomposition}
\label{subsec:rb_phase_bounds}
The main obstacle in solving \eqref{prob:worst_case_information_maximization} lies in constraints \eqref{prob:sinr_constraint} and \eqref{prob:worst_case_information_constraint}. The two semi-infinite constraints share a nonconvex
dependence on the position error.  Specifically, in \eqref{prob:sinr_constraint}, 
\(\mathbf h_k\)  contains factors 
\(e^{j\kappa_{k,\ell}\Delta p_n}\), whereas \(\mathcal I_\theta\) in
\eqref{prob:worst_case_information_constraint} depends on \(\mathbf u\) and
\(\mathbf d\) through \(e^{j\kappa_{\rm t}\Delta p_n}\) and pairwise terms
proportional to
\((p_n-p_m+\Delta p_n-\Delta p_m)
e^{j\kappa_{\rm t}(\Delta p_n+\Delta p_m)}\).  The worst cases over
\(\mathcal E_{\rm p}(\varepsilon)\) therefore cannot be eliminated directly.
 To obtain a form suitable for robust reformulation, we
first consider a generic exponential with a bounded phase argument, say $e^{jz}$ with \(|z|\le x\). 
We use the following lemma to represent this uncertain exponential by an
affine function with a uniformly bounded error.
%, providing a common starting point for reformulating both robust constraints.
%
% The following lemma supplies this building
%block.
%Assumption~\ref{ass:small_position_error} ensures phase radii below \(\pi/2\)
%and \(\pi\) for the individual and pairwise responses, respectively, allowing
%the same decomposition to handle both robust constraints.
\begin{lemma}
\label{lem:rb_tight_phase_remainder}
For a phase radius \(0\le x<\pi\), let
\(c(x)\triangleq(1+\cos x)/2\),
\(a(x)\triangleq(1-\cos x)/2\), and
\(b(x)\triangleq\sin x/x\).
For every real \(z\) satisfying \(|z|\le x\), the phase response admits the
following  decomposition:
\begin{equation}
e^{jz}
 =c(x)+jb(x)z+r_x(z),%
\label{eq:rb_affine_phase_response}%
\end{equation}%
where $ |r_x(z)|\le a(x)$.
\end{lemma}%
The proof is provided in Appendix~\ref{app:proof_tight_phase_remainder}. With Lemma~\ref{lem:rb_tight_phase_remainder}, the nonlinear
phase response reduces to an affine dependence on the
position error \(\Delta\mathbf p\). Next, we use this decomposition to
reformulate the robust communication constraint.

\subsection{Robust Communication Constraint}
\label{subsec:rb_comm_ellipsoids}

Let 
\(\mathbf S_k(\mathbf W)\triangleq
\mathbf w_k\mathbf w_k^{\mathrm H}-
\Gamma_k\sum_{i\ne k}\mathbf w_i\mathbf w_i^{\mathrm H}\),  \eqref{prob:sinr_constraint} can be equivalently written as
\begin{equation}
\mathbf h_k^{\mathrm H}(\tilde{\mathbf p})
\mathbf S_k(\mathbf W)
\mathbf h_k(\tilde{\mathbf p})
\ge \Gamma_k\sigma_k^2,\,\,\,\, \forall\Delta\mathbf p\in\mathcal E_{\rm p}(\varepsilon).
\label{eq:rb_comm_quadratic_sinr}
\end{equation}
The left-hand side of
\eqref{eq:rb_comm_quadratic_sinr} is a nonconvex trigonometric function of
\(\Delta\mathbf p\), and its minimum over \(\mathcal E_{\rm p}(\varepsilon)\) does not
admit a direct tractable finite-dimensional representation.
For path \(\ell\) of CU \(k\), define the corresponding spatial
wavenumber as
\(\kappa_{k,\ell}\triangleq\kappa\cos\theta_{k,\ell}\).
Applying Lemma~\ref{lem:rb_tight_phase_remainder} directly to
\eqref{eq:comm_channel} gives
\begin{equation}
	\mathbf h_k(\tilde{\mathbf p})
	=\bar{\mathbf h}_k(\mathbf p)
	+\mathbf J_k(\mathbf p)\Delta\mathbf p+\mathbf r_k,
	\label{eq:rb_comm_channel_decomposition}
\end{equation}
where the \(n\)-th entry of the affine channel center is
\begin{equation}
	[\bar{\mathbf h}_k]_n
	\triangleq
	\sum_{\ell=1}^{L_k}
	\alpha_{k,\ell}c(\varphi_{k,\ell})
	e^{j\kappa_{k,\ell}p_n},
	\label{eq:rb_comm_channel_center}
\end{equation}
where 
\(\varphi_{k,\ell}\triangleq|\kappa_{k,\ell}|\varepsilon\), and \(\mathbf J_k\in\mathbb C^{N\times N}\) is diagonal with
\begin{equation}
	[\mathbf J_k]_{n,n}
	\triangleq
	j
	\sum_{\ell=1}^{L_k}
	\alpha_{k,\ell}b(\varphi_{k,\ell})\kappa_{k,\ell}
	e^{j\kappa_{k,\ell}p_n}.
	\label{eq:rb_comm_channel_jacobian}
\end{equation}
Let \(\rho_k\triangleq\sqrt{N}\sum_{\ell=1}^{L_k}
|\alpha_{k,\ell}|a(\varphi_{k,\ell})\), the residual $\mathbf r_k$ then satisfies
\(\|\mathbf r_k\|_2\le\rho_k\).
Although \eqref{eq:rb_comm_channel_decomposition} removes the nonconvex
exponential, \eqref{eq:rb_comm_quadratic_sinr} must still hold over
the remaining uncertainty in \(\mathbf J_k\Delta\mathbf p\) and \(\mathbf r_k\).
We therefore enclose their combined effect within an  ellipsoidal uncertainty set. Since \(\Delta\mathbf p\) is real, the real and
imaginary parts of the first-order perturbation are driven by the same
position error vector
\begin{equation}
\begin{bmatrix}
	\Re\{\mathbf J_k\Delta\mathbf p\}\\
	\Im\{\mathbf J_k\Delta\mathbf p\}
\end{bmatrix}
=
\begin{bmatrix}
	\Re\{\mathbf J_k\}\\
	\Im\{\mathbf J_k\}
\end{bmatrix}
\Delta\mathbf p.
\end{equation}
The generating matrix of the minimum-trace outer ellipsoid of the stacked real representation of \(\mathbf J_k(\mathbf p)\Delta\mathbf p\) can then be written as \cite{lorenz2005robust} 
\begin{equation}
	\mathbf R_k(\mathbf p)
	\triangleq
	\varepsilon
	\sqrt{\operatorname{tr}(\boldsymbol\Omega_k)}\,
	\begin{bmatrix}
		\Re\{\mathbf J_k(\boldsymbol\Omega_k^\dagger)^{1/2}\}\\
		\Im\{\mathbf J_k(\boldsymbol\Omega_k^\dagger)^{1/2}\}
	\end{bmatrix}.
	\label{eq:rb_comm_real_first_order_shape}
\end{equation}
where 
\(\boldsymbol\Omega_k\triangleq
\operatorname{diag}(|[\mathbf J_k]_{1,1}|,\ldots,
|[\mathbf J_k]_{N,N}|)\). Furthermore,  combining this ellipsoid with the radius-\(\rho_k\) residual ball gives
the complex generating matrix of the outer ellipsoid
\begin{equation}
	\begin{aligned}
		\mathbf E_k
		\triangleq{}&
		\boldsymbol\Pi\times
		\left[
		(1+\nu_k)\mathbf R_k\mathbf R_k^{\mathrm{T}}
		+(1+\nu_k^{-1})\rho_k^2\mathbf I_{2N}
		\right]^{\frac{1}{2}}.
	\end{aligned}
	\label{eq:rb_comm_shape_matrix}
\end{equation}
where \(\boldsymbol\Pi=[\mathbf I_N,j\mathbf I_N]\) recombines the stacked real  components into complex perturbation, and \(\nu_k\triangleq\sqrt{2N}\rho_k/\|\mathbf R_k\|_{\rm F}\) is the optimal combining factor according to the minimum-trace rule. 
Accordingly, define the ellipsoidal channel uncertainty set
\begin{equation}
\mathcal H_k(\mathbf p)
\triangleq
\left\{
\bar{\mathbf h}_k+\mathbf E_k\widehat{\mathbf z}_k:
\widehat{\mathbf z}_k\in\mathbb R^{2N},\ 
\|\widehat{\mathbf z}_k\|_2\le1
\right\}.
\label{eq:rb_comm_channel_ellipsoid}
\end{equation}
Every channel vector 
\(\mathbf h_k(\mathbf p+\Delta\mathbf p)\) for all $\Delta\mathbf p\in\mathcal E_{\rm p}(\varepsilon) $ belongs to
\(\mathcal H_k(\mathbf p)\). 
A sufficient condition for
\eqref{eq:rb_comm_quadratic_sinr} is therefore
\begin{equation}
\mathbf h_k^{\mathrm H}
\mathbf S_k(\mathbf W)
\mathbf h_k
\ge\Gamma_k\sigma_k^2,\,\,\,\, \forall\mathbf h_k\in\mathcal H_k(\mathbf p).\label{renmeib}
\end{equation}
Augment the complex generating matrix by
\(\mathbf T_k\triangleq
[\mathbf E_k,\bar{\mathbf h}_k]
\). So that for every \(\mathbf h_k\in\mathcal H_k(\mathbf p)\), one can write
\(\mathbf h_k=\mathbf T_k
[\widehat{\mathbf z}_k^{\rm T},1]^{\rm T}\).
According to S-lemma \cite{vucic2009robust}, the constraint \eqref{renmeib} holds if and only
if there exists \(\mu_k\geq0\) such that
\begin{equation}
\begin{aligned}
\mathbf M_k
\triangleq{}
\Re\!\left\{
\mathbf T_k^{\mathrm H}\mathbf S_k(\mathbf W)\mathbf T_k
\right\}
+\mu_k\mathbf C-\Gamma_k\sigma_k^2
\mathbf e_{2N+1}\mathbf e_{2N+1}^{\mathrm T}
\succeq\mathbf0.
\end{aligned}
\label{eq:rb_robust_comm_lmi}
\end{equation}
where
\(\mathbf C\triangleq\operatorname{blkdiag}(\mathbf I_{2N},-1)\).
Through \eqref{eq:rb_robust_comm_lmi}, the original semi-infinite SIN  requirement \eqref{prob:sinr_constraint} reduces to
a finite-dimensional positive-semidefinite
constraint. 
\subsection{Robust Sensing Constraint}
\label{subsec:rb_sensing_constraint_reformulation}

With \eqref{eq:crb_theta}, constraint
\eqref{prob:worst_case_information_constraint} can be written as
\begin{equation}
\begin{aligned}
\mathcal I_\theta(\mathbf W,\tilde{\mathbf p})=
\eta_{\rm r}\|\mathbf u\|_2^2
+M\mathcal R(\mathbf W,\tilde{\mathbf p})
\frac{\|\mathbf d\|_2^2}{\|\mathbf u\|_2^2}
\ge\tau,\;
\forall\Delta\mathbf p\in\mathcal E_{\rm p}(\varepsilon).
\end{aligned}
\label{eq:rb_expanded_sensing_constraint}
\end{equation}
where 
\(\mathcal R(\mathbf W,\tilde{\mathbf p})\triangleq
\frac{
\tilde{\mathbf p}^{\rm T}\mathbf J_{\rm e}
(\mathbf J_{\rm e}+\mathbf J_{\rm d})^{-1}
\mathbf J_{\rm d}\tilde{\mathbf p}}
{\tilde{\mathbf p}^{\rm T}\mathbf J_{\rm e}\tilde{\mathbf p}}\).
As can be observed, \eqref{eq:rb_expanded_sensing_constraint} has a complicated
form and is therefore difficult to handle directly.
We next construct a lower bound on the left-hand side for a more tractable form. We first handle \(\mathcal R(\mathbf W,\tilde{\mathbf p})\), which
represents the fraction of transmit-aperture angle information retained under imperfect APV knowledge. 
Since \(\mathcal R(\mathbf W,\tilde{\mathbf p})\) is governed by the
balance between \(\mathbf J_{\rm d}\) and \(\mathbf J_{\rm e}\), we characterize
this balance by upper-bounding  \(\mathbf J_{\rm e}\) with
\(\mathbf J_{\rm e}
\preceq
\zeta\mathbf J_{\rm d}(\tilde{\mathbf p})\), 
where \(\zeta\ge0\) is introduced as an auxiliary variable. This 
implies
\begin{equation}
	\mathbf J_{\rm e}
	(\mathbf J_{\rm e}+\mathbf J_{\rm d})^{-1}
	\mathbf J_{\rm d}
	\succeq
	\frac{1}{1+\zeta}\mathbf J_{\rm e},
	\label{eq:rb_retained_information_matrix_bound}
\end{equation}
which further gives $\mathcal R(\mathbf W,\tilde{\mathbf p})$ a lower bound as 
\(\mathcal R(\mathbf W,\tilde{\mathbf p})
\ge\frac{1}{1+\zeta}.\)
To account for the unknown position error in $\mathbf J_{\rm d}$, for the \(n\)-th MA, define its worst-case pilot information as
\(\underline g_{\rm d}(p_n)\triangleq
\min_{|\Delta p_n|\le\varepsilon}
g_{\rm d}(\tilde{p}_n)\). Define $	\underline{\mathbf J}_{\rm d}(\mathbf p)\triangleq	\operatorname{diag}\!\left(
\underline g_{\rm d}(p_1),\ldots,
\underline g_{\rm d}(p_N)
\right) $ as the worst-case pilot information matrix for the MA position errors. Since \(\underline{\mathbf J}_{\rm d}(\mathbf p)\preceq
\mathbf J_{\rm d}(\tilde{\mathbf p})\),  \eqref{eq:rb_retained_information_matrix_bound} can  be ensured by
\begin{equation}
{\mathbf J}_{\rm e}
	\preceq\zeta
\underline{\mathbf J}_{\rm d}({\mathbf p}),
	\label{eq:rb_echo_pilot_domination}
\end{equation}
In particular, $\underline g_{\rm d}$ can be efficiently evaluated through a one-dimensional bounded search. Using 
the definition of \(\mathbf J_{\rm e}\) in
\eqref{eq:echo_position_information} and the fact that
\(\mathbf0\preceq\mathbf P_{\mathbf u}^{\perp}\preceq\mathbf I_K\), \eqref{eq:rb_echo_pilot_domination} can be guaranteed by the following positive-semidefinite constraint through the Schur complement
\begin{equation}
	\mathbf M_{\rm s}(\mathbf W,\mathbf p,\zeta)
	\triangleq
	\begin{bmatrix}
		\zeta\mathbf I_K
		&
		\sqrt{\gamma_{\rm s}M}\,|\kappa_{\rm t}|\mathbf W^{\rm H}\\
		\sqrt{\gamma_{\rm s}M}\,|\kappa_{\rm t}|\mathbf W
		&
		\underline{\mathbf J}_{\rm d}(\mathbf p)
	\end{bmatrix}
	\succeq\mathbf0.
	\label{eq:rb_spectral_domination_lmi}
\end{equation}
%Indeed, \eqref{eq:rb_spectral_domination_lmi} ensures
%\(\mathbf J_{\rm e}
%\preceq\zeta\underline{\mathbf J}_{\rm d}(\mathbf p)
%\preceq\zeta\mathbf J_{\rm d}(\tilde{\mathbf p})\)
%for every admissible position error.  Hence, every feasible \(\zeta\)
%certifies the uniform lower bound
%\(\mathcal R(\mathbf W,\tilde{\mathbf p})\ge1/(1+\zeta)\), with a smaller
%\(\zeta\) providing a tighter bound.
%Hence, \(\mathcal R(\mathbf W,\tilde{\mathbf p})\ge1/(1+\zeta)\), and a smaller feasible
%\(\zeta\) retains more transmit-aperture information.
%\begin{remark}
%If \(\kappa_{\rm t}=0\), then \(\mathbf G=\mathbf0\), no
%position--angle coupling remains, and one sets \(\zeta=0\) directly; the
%transmit-aperture fraction reduces to
%\(\|\mathbf d\|_2^2/\|\mathbf u\|_2^2\).
%\end{remark}
%Combining \eqref{eq:rb_echo_pilot_domination} with Proposition~\ref{prop:rb_exact_aperture_decomposition} now gives a lower bound in terms
%of \(\mathbf u\) and \(\mathbf d\).  
Therefore, after lower-bounding
\(\mathcal R(\mathbf W,\tilde{\mathbf p})\), constraint
\eqref{eq:rb_expanded_sensing_constraint} is safely enforced by
\eqref{eq:rb_spectral_domination_lmi} together with
\begin{equation}
	\eta_{\rm r}\|\mathbf u\|_2^2
	+
	\frac{M}{1+\zeta}\frac{\|\mathbf d\|_2^2}
	{\|\mathbf u\|_2^2}
	\ge\tau,\,\,\,\,
	\forall\Delta\mathbf p\in\mathcal E_{\rm p}(\varepsilon).
	\label{eq:rb_sensing_after_R_bound}
\end{equation}
The constraint \eqref{eq:rb_sensing_after_R_bound} remains intractable and contains a quadratic receive-aperture term
and a fractional transmit-aperture term. We express the former using
the standard variational representation
and handle the latter with the quadratic transform \cite{shen2018fractional}. Accordingly, introducing two
auxiliary variables
\(\mathbf q\in\mathbb C^K\) and
\(\mathbf t\in\mathbb C^{\binom{K}{2}}\) yields the equivalent transform
\begin{equation}
\begin{aligned}
\eta_{\rm r}\|\mathbf u\|_2^2+
\frac{M}{1+\zeta}\frac{\|\mathbf d\|_2^2}{\|\mathbf u\|_2^2}
&=\max_{\mathbf q,\mathbf t}
\Big\{
2\eta_{\rm r}\Re\{\mathbf q^{\rm H}\mathbf u\}
-\eta_{\rm r}\|\mathbf q\|_2^2\\
&+2M\Re\{\mathbf t^{\rm H}\mathbf d\}
-(1+\zeta)M\|\mathbf t\|_2^2\|\mathbf u\|_2^2
\Big\}.
\end{aligned}
\label{eq:rb_joint_information_representation}
\end{equation}
In order to obtain a 
finite-dimensional formulation, it remains to lower-bound the left-hand side of
\eqref{eq:rb_sensing_after_R_bound} over the 
position error set. We treat
\(\mathbf q\) and \(\mathbf t\) as a pair of auxiliary variables whose
values are common to all admissible position errors. By the minimax inequality, any such common pair provides a lower bound on the worst-case value of the
left-hand side of \eqref{eq:rb_sensing_after_R_bound} over
\(\mathcal E_{\rm p}(\varepsilon)\). 
For a fixed candidate pair $(\mathbf q,\mathbf t)$, we first handle the
last negative term in \eqref{eq:rb_joint_information_representation} by
constructing an upper bound on $\|\mathbf u\|_2^2$. Let
$\varphi_{\rm t}\triangleq
|\kappa_{\rm t}|\varepsilon<\pi/2$.
For every $|z|\le\varphi_{\rm t}$,
$|e^{jz}-\cos\varphi_{\rm t}|\le\sin\varphi_{\rm t}$.
Applying this further to entries of \(\mathbf a_{\rm t}(\mathbf p+\Delta\mathbf p)\)
gives
\begin{equation}
	\left\|
	\mathbf a_{\rm t}(\mathbf p+\Delta\mathbf p)
	-\cos\varphi_{\rm t}\mathbf a_{\rm t}(\mathbf p)
	\right\|_2^2
	\le N\sin^2\varphi_{\rm t}.
	\label{eq:rb_steering_enclosing_ball}
\end{equation}
 Define
	$\bar{\mathbf u}(\mathbf W,\mathbf p)
	\triangleq
	\cos\varphi_{\rm t}\mathbf W^{\rm H}
	\mathbf a_{\rm t}(\mathbf p)$.
Then, $\mathbf u$ can be written as
\begin{equation}
	\mathbf u
	=
	\bar{\mathbf u}(\mathbf W,\mathbf p)
	+\mathbf W^{\rm H}
	\left[
	\mathbf a_{\rm t}(\mathbf p+\Delta\mathbf p)
	-\cos\varphi_{\rm t}\mathbf a_{\rm t}(\mathbf p)
	\right], \label{eq:rb_echo_ball_mapping}
\end{equation}
where the bracketed term lies in the ball specified by
\eqref{eq:rb_steering_enclosing_ball}. Hence, 
$\mathbf u$ lies in the image of this ball under $\mathbf W^{\rm H}$,
whose center is $\bar{\mathbf u}(\mathbf W,\mathbf p)$.
We further introduce an
auxiliary variable $\overline U\ge0$ to uniformly upper-bound $	\|\mathbf u\|_2^2$ under every admissible position error, namely 
\begin{equation}
	\|\mathbf u\|_2^2\le\overline U,
\,\,\,\,
	\forall\,\Delta\mathbf p\in\mathcal E_{\rm p}(\varepsilon).
	\label{eq:rb_echo_energy_upper_bound}
\end{equation}
%Substituting \eqref{eq:rb_echo_ball_mapping} into
%\eqref{eq:rb_echo_energy_upper_bound} gives
%\begin{equation}
%	\left\|
%	\bar{\mathbf u}(\mathbf W,\mathbf p)
%	+\mathbf W^{\rm H}
%	\left[
%	\mathbf a_{\rm t}(\mathbf p+\Delta\mathbf p)
%	-\cos\varphi_{\rm t}\mathbf a_{\rm t}(\mathbf p)
%	\right]
%	\right\|_2^2
%	\le\overline U.
%	\label{eq:rb_echo_energy_quadratic_bound}
%\end{equation}
Combining \eqref{eq:rb_steering_enclosing_ball} and \eqref{eq:rb_echo_ball_mapping}, the S-lemma gives the following finite-dimensional sufficient condition
for \eqref{eq:rb_echo_energy_upper_bound}:
\begin{equation}
	\mathbf M_u(\mathbf W,\mathbf p,\overline U,\lambda_u)
	\triangleq
	\begin{bmatrix}
		\overline U-\lambda_uN\sin^2\varphi_{\rm t}
		& \mathbf0
		& \bar{\mathbf u}^{\rm H}\\
		\mathbf0
		& \lambda_u\mathbf I_N
		& \mathbf W\\
		\bar{\mathbf u}
		& \mathbf W^{\rm H}
		& \mathbf I_K
	\end{bmatrix}
	\succeq\mathbf0.
	\label{eq:rb_echo_trust_region_lmi}
\end{equation}
where $\lambda_u\ge0$ is an auxiliary variable. 
We next jointly lower-bound the two
positive terms in
\eqref{eq:rb_joint_information_representation} as follows.
\begin{proposition}
\label{prop:rb_positive_response_lower_bound}
For any \(\Delta\mathbf p\in\mathcal E_{\rm p}(\varepsilon)\),
\begin{equation}
2\eta_{\rm r}\Re\{\mathbf q^{\rm H}\mathbf u\}
+2M\Re\{\mathbf t^{\rm H}\mathbf d\}
\ge \underline\Phi(\mathbf W,\mathbf p),
\label{eq:rb_positive_response_lower_bound}
\end{equation}
where \(\underline\Phi(\mathbf W,\mathbf p)\) is given by
\begin{equation}
\begin{aligned}
\underline\Phi(\mathbf W,\mathbf p)&\triangleq 2\Big[
\eta_{\rm r}c(\varphi_{\rm t})
\Re\{\mathbf1_N^{\rm T}\mathbf z_u\}
+M\Re\{\mathbf t^{\rm H}\bar{\mathbf d}\}-\varepsilon\|\mathbf g\|_1\\
&
-\eta_{\rm r}a(\varphi_{\rm t})\|\mathbf z_u\|_1
-M\boldsymbol\rho_d^{\rm T}|\mathbf C_d^{\rm H}\mathbf t|
\Big].
\end{aligned}
\label{eq:rb_phi}
\end{equation}
In \eqref{eq:rb_phi}, 
\(\mathbf z_u\triangleq(\mathbf W\mathbf q)^*
\odot\mathbf a_{\rm t}(\mathbf p)\). The matrix
\(\mathbf C_d(\mathbf W,\mathbf p)\in
\mathbb C^{\binom{K}{2}\times\binom{N}{2}}\) has entries
$[\mathbf C_d]_{(k,\ell),(n,m)}
\triangleq
-j\beta_{\rm t}[\mathbf a_{\rm t}(\mathbf p)]_n
[\mathbf a_{\rm t}(\mathbf p)]_m
\left(w_{n,k}w_{m,\ell}-w_{n,\ell}w_{m,k}\right)^*$ for index \(1\le k<\ell\le K\) and \(1\le n<m\le N\). $[\boldsymbol\rho_d]_{(n,m)}
\triangleq
(p_m-p_n+2\varepsilon)a(2\varphi_{\rm t})
+b(2\varphi_{\rm t})|\kappa_{\rm t}|
\varepsilon^2 $, \(1\le n<m\le N\). Define the real vector
\(\mathbf g\triangleq M\Re\{\mathbf A_{\rm d}^{\rm H}\mathbf t\}
-\eta_{\rm r}b(\varphi_{\rm t})\kappa_{\rm t}
\Im\{\mathbf z_u\}\). Furthermore, 
\begin{subequations}
\begin{equation}
	\bar{\mathbf d}
	\triangleq
	c(2\varphi_{\rm t})\mathbf C_d
	\mathbf B_{\Delta}\mathbf p.
	\label{eq:rb_pair_affine_center}
\end{equation}
\begin{equation}
	\mathbf A_{\rm d}
	\triangleq
	\mathbf C_d\Big[
	c(2\varphi_{\rm t})\mathbf B_{\Delta}
	+jb(2\varphi_{\rm t})\kappa_{\rm t}
	\operatorname{diag}(\mathbf B_{\Delta}\mathbf p)
	\mathbf B_{\Sigma}
	\Big].
	\label{eq:rb_pair_affine_jacobian}
\end{equation}
\end{subequations}
where \(\mathbf B_{\Delta},\mathbf B_{\Sigma}
\in\mathbb R^{\binom{N}{2}\times N}\) have rows
\(\mathbf e_n^{\rm T}-\mathbf e_m^{\rm T}\) and
\(\mathbf e_n^{\rm T}+\mathbf e_m^{\rm T}\), respectively, for every
\(1\le n<m\le N\).
\end{proposition}
The proof 
is provided in Appendix~\ref{app:proof_positive_response_lower_bound}.
% And the definition fo $a(\cdot)$, $b(\cdot)$ and $c(\cdot)$ follow from Lemma~\ref{lem:rb_tight_phase_remainder}. 
As a consequence,
for the common auxiliary pair $(\mathbf q,\mathbf t)$, combining Proposition~\ref{prop:rb_positive_response_lower_bound} and the upper bound
$\|\mathbf u\|_2^2\le\overline U$ yields a sufficient lower
bound on the left-hand side of
\eqref{eq:rb_sensing_after_R_bound}, which is
\begin{equation}
	\underline\Phi(\mathbf W,\mathbf p)
	-\eta_{\rm r}\|\mathbf q\|_2^2
	-M(1+\zeta)\|\mathbf t\|_2^2\overline U
	\ge\tau.
	\label{eq:rb_robust_sensing_information}
\end{equation}
Consequently,
\eqref{eq:rb_spectral_domination_lmi},
\eqref{eq:rb_echo_trust_region_lmi}, and
\eqref{eq:rb_robust_sensing_information}, together with the
nonnegativity of the associated auxiliary variables, constitute a
finite-dimensional sufficient reformulation of the original semi-infinite
sensing constraint
\eqref{eq:rb_expanded_sensing_constraint}.
%These constraints guarantee
%$\mathcal I_\theta(\mathbf W,\mathbf p+\Delta\mathbf p)\ge\tau$
%for every $\Delta\mathbf p\in\mathcal E_{\rm p}(\varepsilon)$ and, consequently,
%a worst-case angle CRB no greater than
%$1/(\gamma_{\rm s}\tau)$.
\subsection{Alternating Optimization Framework}
\label{subsec:rb_ao_framework}

Collecting the communication and sensing reformulations, the
semi-infinite problem \eqref{prob:worst_case_information_maximization} is safely approximated by the following
finite-dimensional problem:
\begin{subequations}
\label{prob:rb_finite_dimensional_problem}
\begin{align}
\max_{\substack{
\mathbf W,\mathbf p,\tau,\mathbf q,\mathbf t,\\
\overline U,\zeta,\lambda_u,\{\mu_k\}}}
\quad &\tau
\label{prob:rb_finite_dimensional_objective}\\
\mathrm{s.t.}\quad
&\eqref{prob:power_constraint},\
\eqref{prob:nominal_apv_constraint}, \eqref{eq:rb_spectral_domination_lmi}, \eqref{eq:rb_echo_trust_region_lmi},
\eqref{eq:rb_robust_sensing_information},
\label{prob:rb_finite_dimensional_design_constraints}\\
&\eqref{eq:rb_robust_comm_lmi},
\,\,\,\, k\in\mathcal K,
\label{prob:rb_finite_dimensional_comm_constraint}\\
&\overline U,
\zeta,\lambda_u,\mu_k\ge0,\,\,\,\, k\in\mathcal K.
\label{prob:rb_finite_dimensional_variable_constraints}
\end{align}
\end{subequations}
Problem \eqref{prob:rb_finite_dimensional_problem} remains nonconvex.  We therefore solve it 
by alternating among the auxiliary, beamforming, and APV blocks.
%At each outer iteration, the auxiliary variables and certificates are first
%tightened at the current beamformer and APV.  A cyclic beam-coordinate sweep is
%then performed, followed by an inner majorization--minimization (MM) loop for
%the APV update.
\subsubsection{Auxiliary-variable update}
\label{subsec:rb_auxiliary_update}

At the beginning of outer iteration \(r\), we first update the auxiliary block at fixed
\((\mathbf W^{(r)},\mathbf p^{(r)})\). According to
\eqref{eq:rb_spectral_domination_lmi}, the  optimal value of
 \(\zeta\) is available in closed form as
\(\zeta_\star(\mathbf W,\mathbf p)
\triangleq\gamma_{\rm s}M\kappa_{\rm t}^2
\lambda_{\max}(\mathbf W^{\rm H}
\underline{\mathbf J}_{\rm d}^{-1}(\mathbf p)\mathbf W)\).  The
remaining auxiliary variables are updated as
\begin{subequations}
\label{eq:rb_joint_auxiliary_update}
\begin{align}
\big(\overline U^{(r)},\lambda_u^{(r)}\big)
&\in\arg\min_{\overline U,\lambda_u\ge0}\ \overline U\notag\\
&\quad\mathrm{s.t.}\quad
\mathbf M_u(\mathbf W^{(r)},\mathbf p^{(r)},
\overline U,\lambda_u)\succeq\mathbf0,
\label{eq:rb_echo_energy_update}\\
(\mathbf q^{(r)},\mathbf t^{(r)})
&\in\arg\max_{\mathbf q,\mathbf t}\Big\{
\underline\Phi(\mathbf W^{(r)},\mathbf p^{(r)};
\mathbf q,\mathbf t)\notag\\
&-\eta_{\rm r}\|\mathbf q\|_2^2
-M\|\mathbf t\|_2^2\overline U^{(r)}(1+\zeta^{(r)})
\Big\}.
\label{eq:rb_auxiliary_vector_update}
\end{align}
\end{subequations}
Both optimization problems in \eqref{eq:rb_joint_auxiliary_update} are convex.
%The auxiliary vectors
%\(\mathbf q^{(r)}\) and \(\mathbf t^{(r)}\) are held fixed throughout the
%subsequent beamforming and APV updates, whereas \(\overline U\), \(\zeta\),
%\(\lambda_u\), and \(\{\mu_k\}\) are reoptimized jointly with the
%corresponding design variables.

The scalar sensing constraint \eqref{eq:rb_robust_sensing_information}
contains the bilinear product \(\overline U\zeta\) in
\(\overline U(1+\zeta)\).  Jointly optimizing \(\overline U\) and \(\zeta\)
therefore prevents the subsequent beamforming and APV subproblems from being
convex.  Given their current reference values
\((\overline U_0,\zeta_0)\), define the common upper model
\begin{equation}
\begin{aligned}
\overline D_0(\overline U,\zeta)
\triangleq\overline U+\frac12\left[
\frac{\zeta_0}{\overline U_0}\overline U^2
+\frac{\overline U_0}{\zeta_0}\zeta^2
\right].
\end{aligned}
\label{eq:rb_denominator_upper_bound}
\end{equation}
By weighted Young's inequality, \(\overline D_0\) is a convex,
first-order-tight upper bound on \(\overline U(1+\zeta)\) at
\((\overline U_0,\zeta_0)\).  Accordingly, in the subsequent beamforming
and APV updates, \eqref{eq:rb_robust_sensing_information} is replaced by
the following inner constraint:
\begin{equation}
\underline\Phi(\mathbf W,\mathbf p)
-\eta_{\rm r}\|\mathbf q\|_2^2
-M\|\mathbf t\|_2^2\overline D_0(\overline U,\zeta)
\ge\tau.
\label{eq:rb_inner_sensing_information}
\end{equation}
%When \(\kappa_{\rm t}=0\), set \(\zeta=0\) and use the linear denominator
%\(\overline U\) directly, without \(\overline D_0\).

\subsubsection{Beamforming update}
\label{subsec:rb_beamforming_update}
The beamforming optimizer updates 
\(\mathbf w_1,\ldots,\mathbf w_K\) sequentially. With
\((\mathbf p,\mathbf q,\mathbf t)\) and all nonactive columns fixed at their latest values, constraint
\eqref{prob:power_constraint}, \eqref{eq:rb_spectral_domination_lmi},
\eqref{eq:rb_echo_trust_region_lmi} and
\eqref{eq:rb_inner_sensing_information} are convex in
\(\mathbf w_j\).  Before solving
the \(j\)-th coordinate problem, the reference values
\((\overline U_0,\zeta_0)\) in \eqref{eq:rb_denominator_upper_bound} are set to
their current feasible values.
It remains to obtain convex forms for \eqref{prob:rb_finite_dimensional_comm_constraint}.  The desired covariance
of CU \(j\) admits the affine lower bound
$\underline{\mathbf Q}_j^{(r)}(\mathbf w_j)
\triangleq{}
\mathbf w_j^{(r)}\mathbf w_j^{\mathrm H}
+\mathbf w_j(\mathbf w_j^{(r)})^{\mathrm H}
-\mathbf w_j^{(r)}(\mathbf w_j^{(r)})^{\mathrm H}
\preceq\mathbf w_j\mathbf w_j^{\mathrm H}$.
Consequently,  the \(j\)-th QoS constraint is represented by
\begin{equation}
\begin{split}
&\Re\!\left\{\mathbf T_j^{\mathrm H}
\underline{\mathbf Q}_j^{(r)}(\mathbf w_j)\mathbf T_j\right\}
+\mu_j\mathbf C-\Gamma_j\sigma_j^2\mathbf e_{2N+1}\mathbf e_{2N+1}^{\mathrm T}\\&-\Gamma_j\Re\!\left\{\mathbf T_j^{\mathrm H}
\left(\sum_{\ell\ne j}\mathbf w_\ell
\mathbf w_\ell^{\mathrm H}\right)\mathbf T_j\right\}
\succeq\mathbf0.
\end{split}
\label{eq:rb_coordinate_desired_comm_lmi}
\end{equation}
For CU \(k\ne j\), collect the covariance independent of the active
beam \(\mathbf w_j\) as
$\mathbf S_{k,j}
\triangleq
\mathbf w_k\mathbf w_k^{\mathrm H}
-\Gamma_k\!\sum_{\ell\notin\{k,j\}}
\mathbf w_\ell\mathbf w_\ell^{\mathrm H}$.  Further define
\(\mathbf F_{k,j}(\mu_k)\triangleq
\Re\{\mathbf T_k^{\mathrm H}\mathbf S_{k,j}\mathbf T_k\}
+\mu_k\mathbf C
-\Gamma_k\sigma_k^2\mathbf e_{2N+1}\mathbf e_{2N+1}^{\mathrm T}\) and
\(\mathbf V_{k,j}(\mathbf w_j)\triangleq
\bigl[\Re\{\mathbf w_j^{\mathrm H}\mathbf T_k\}^{\mathrm T},
\Im\{\mathbf w_j^{\mathrm H}\mathbf T_k\}^{\mathrm T}\bigr]^{\mathrm T}\).
For CU \(k\ne j\), the active beam \(\mathbf w_j\) enters
\eqref{eq:rb_robust_comm_lmi} only through the interference term
\(-\Gamma_k\Re\{\mathbf T_k^{\mathrm H}\mathbf w_j
\mathbf w_j^{\mathrm H}\mathbf T_k\}\).
Since
$
\Re\{\mathbf T_k^{\mathrm H}\mathbf w_j
\mathbf w_j^{\mathrm H}\mathbf T_k\}
=
\mathbf V_{k,j}^{\mathrm T}(\mathbf w_j)
\mathbf V_{k,j}(\mathbf w_j)$,
the \(k\)-th robust QoS constraint is equivalently expressed, through
the Schur complement, as
\begin{equation}
	\begin{bmatrix}
		\mathbf F_{k,j}(\mu_k)
		&\sqrt{\Gamma_k}\mathbf V_{k,j}^{\mathrm T}(\mathbf w_j)\\
		\sqrt{\Gamma_k}\mathbf V_{k,j}(\mathbf w_j)&\mathbf I_2
	\end{bmatrix}
	\succeq\mathbf0,
	\quad k\ne j.
	\label{eq:rb_coordinate_interference_lmi}
\end{equation}
Finally, the complete update of beam \(j\) is given by
\begin{subequations}
\label{prob:rb_coordinate_beam_update}
\begin{align}
\max_{\substack{
\mathbf w_j,\tau,\overline U,\zeta,\lambda_u,
\{\mu_k\}}}
\quad
&\tau\\
\mathrm{s.t.}\quad
&\eqref{prob:power_constraint},\eqref{eq:rb_spectral_domination_lmi}, \eqref{eq:rb_echo_trust_region_lmi},
\eqref{eq:rb_inner_sensing_information}, \eqref{eq:rb_coordinate_desired_comm_lmi}, \eqref{prob:rb_finite_dimensional_variable_constraints}.\\
&\eqref{eq:rb_coordinate_interference_lmi},\,\,\,\, k\in\mathcal{K}, k\neq j,
\end{align}
\end{subequations}
All constraints in \eqref{prob:rb_coordinate_beam_update} are convex in the
active block, so it can be solved by standard conic solvers.  After one
cyclic sweep for all $\mathbf{w}_j$, the resulting matrix is denoted by \(\mathbf W^{(r+1)}\).

\subsubsection{APV update}
\label{subsec:rb_apv_update}
With
\((\mathbf W,\mathbf q,\mathbf t)\) fixed at their latest values,  initialize
\(\mathbf p^{(r,0)}=\mathbf p^{(r)}\).
We use an analytical quadratic lower-model operator for the APV-dependent
quantities.  For an \(m\times m\) Hermitian matrix function
\(\mathbf F(\mathbf p)\), define
\begin{equation}
\begin{aligned}
\mathcal L^{(r,i)}[\mathbf F]
(\mathbf p)
&\triangleq{}
\mathbf F(\mathbf p^{(r,i)})
+\mathrm d\mathbf F(\mathbf p^{(r,i)})
[\mathbf p-\mathbf p^{(r,i)}]\\
&-\frac12
(\mathbf p-\mathbf p^{(r,i)})^{\mathrm T}
\mathbf H_{\mathbf F}^{(r,i)}
(\mathbf p-\mathbf p^{(r,i)})\mathbf I_m,
\end{aligned}
\label{eq:rb_paired_apv_models}
\end{equation}
where \(\mathrm d\mathbf F(\mathbf p)[\cdot]\) is the directional
derivative, and \(\mathbf H_{\mathbf F}^{(r,i)}\) is a diagonal curvature
matrix chosen so that
\(\mathcal L^{(r,i)}[\mathbf F](\mathbf p)\preceq\mathbf F(\mathbf p)\)
on the current smooth branch, with value and first-order equality at the
expansion point, as follows from the standard quadratic MM
\cite{sun2017mm}. The required first-order differentials and the generic curvature construction are summarized in Appendix~\ref{app:analytic_apv_models}.
%, which also gives the evaluation of the
%one-dimensional minima in the definition of
%\(\underline{\mathbf J}_{\rm d}(\mathbf p)\). 
Notably, the parameters \(\boldsymbol\Omega_k\) and \(\nu_k\) are
evaluated at \(\mathbf p^{(r,i)}\) and held fixed in the current subproblem, while the resulting ellipsoid remains a valid outer set over the APV update.
%At the current APV iterate, the reference values in
%\eqref{eq:rb_denominator_upper_bound} are set to the current feasible echo
%upper bound \(\overline U_0\) and
%\(\zeta_0=\zeta_\star(\mathbf W^{(r+1)},\mathbf p^{(r,i)})\). 
The APV subproblem is
given by
\begin{subequations}
\label{prob:rb_apv_update}
\begin{align}
\max_{\substack{
\mathbf p,\tau,\overline U,\zeta,\lambda_u,
\{\mu_k\}}}
\quad
&\tau\\
\mathrm{s.t.}\quad
&\eqref{prob:nominal_apv_constraint}, \eqref{prob:rb_finite_dimensional_variable_constraints},\\
&\mathcal L^{(r,i)}[\mathbf M_{\rm s}(\zeta)](\mathbf p)
\succeq\mathbf0,\\
&\mathcal L^{(r,i)}[\underline\Phi](\mathbf p)
-\eta_{\rm r}\|\mathbf q^{(r)}\|_2^2\notag\\
&\quad
-M\|\mathbf t^{(r)}\|_2^2
\overline D_0(\overline U,\zeta)\ge\tau,\\
&\mathcal L^{(r,i)}[\mathbf M_u(
\overline U,\lambda_u)](\mathbf p)\succeq\mathbf0,\\
&\mathcal L^{(r,i)}[\mathbf M_k(\mu_k)](\mathbf p)
\succeq\mathbf0,
\,\,\,\, k\in\mathcal K.
\end{align}
\end{subequations} 
The solution of \eqref{prob:rb_apv_update} is accepted through a standard
backtracking line search that preserves feasibility and does not decrease
\(\tau\). The inner loop terminates when the relative objective improvement is
below tolerance \(\epsilon_{\rm APV}\).

\subsection{Overall Algorithm and Analysis}
\label{subsec:rb_overall_algorithm}
During the initialization stage of the algorithm, the APV is initialized by uniformly spacing the MAs across the full feasible aperture. While $\mathbf W^{(0)}$ is initialized by a
nominal zero-forcing (ZF) beamformer and
iteratively refining it to satisfy \eqref{eq:rb_robust_comm_lmi} if necessary. 
Algorithm~\ref{alg:rb_ao} summarizes the procedure.  
%Initialize
%\(\mathbf p^{(0)}\) by uniformly spacing the MAs across the feasible region.
%With this APV fixed, a communication-only successive convex approximation (SCA)
%initialized by the nominal-channel zero-forcing beamformer yields \(\mathbf W^{(0)}\) satisfying
%\eqref{eq:rb_robust_comm_lmi} and the power constraint.  Uniformly scale this
%feasible beamformer to use \(P_{\max}\) before the sensing iterations.
\begin{algorithm}[t]
\caption{Single-Sweep Beam-Coordinate Robust AO}
\label{alg:rb_ao}
\begin{algorithmic}[1]
\REQUIRE Initialize 
\((\mathbf W^{(0)},\mathbf p^{(0)})\), tolerances
\(\epsilon_{\rm AO}\) and \(\epsilon_{\rm APV}\).
\STATE Set \(r=0\) and compute \(\tau^{(0)}\).
\REPEAT
\STATE Update \(\zeta^{(r)}\) in closed form and the remaining
auxiliary variables by \eqref{eq:rb_joint_auxiliary_update}.
\FOR{\(j=1,\ldots,K\)}
\STATE Update \(\mathbf w_j\) by solving
\eqref{prob:rb_coordinate_beam_update}.
\ENDFOR
\STATE Set \(\mathbf W^{(r+1)}\leftarrow\mathbf W\).
\STATE With \((\mathbf W, \mathbf q,\mathbf t)\) fixed, update the APV
block by solving \eqref{prob:rb_apv_update} with the inner MM and
backtracking procedure until the improvement is below
\(\epsilon_{\rm APV}\), yielding
\((\mathbf p^{(r+1)},\tau^{(r+1)})\).
\STATE Set \(r\leftarrow r+1\).
\UNTIL{\(|\tau^{(r)}-\tau^{(r-1)}|
\le\epsilon_{\rm AO}\max\{1,\tau^{(r-1)}\}\)}
\ENSURE \((\mathbf W^\star,\mathbf p^\star)=(\mathbf W^{(r)},\mathbf p^{(r)})\).
\end{algorithmic}
\end{algorithm}
The returned value \(\tau^{(r)}\) certifies the worst-case CRB upper bound
\(\sigma_{\rm s}^2/(2T|\alpha_{\rm t}|^2\tau^{(r)})\).
Every accepted block update preserves feasibility and does not
decrease \(\tau\).  Since \(\tau\) is bounded under \(P_{\max}\), the
nondecreasing objective sequence \(\{\tau^{(r)}\}\) converges.  To characterize
the computational cost, let \(I_{\rm AO}\) and \(I_{\rm APV}\) denote the
numbers of outer iterations and average APV inner iterations, respectively.
Each outer iteration performs the closed-form \(\zeta\) update, two auxiliary
convex updates, \(K\) beam-coordinate SDPs, and \(I_{\rm APV}\) APV SDPs.  The
largest LMI has order \(\max\{N+K+1,2N+3\}\).  Excluding initialization, the
overall complexity is
\(\mathcal O\!\left(I_{\rm AO}(\mathcal C_{\rm aux}
+K\mathcal C_{\rm BF}+I_{\rm APV}\mathcal C_{\rm APV})\right)\), where the
three terms denote the respective per-update costs. 

\section{Numerical Results}
\label{sec:numerical_results}

In this section, numerical simulations are carried out to evaluate the proposed
robust MA-ISAC beamforming design. 
\begin{table}[t]
	\caption{Default Simulation Parameters}
	\label{tab:simulation_parameters}
	\centering
	\footnotesize
	\renewcommand{\arraystretch}{1.18}
	\begin{tabular}{p{0.46\linewidth}p{0.42\linewidth}}
		\hline
		Parameter & Value \\
		\hline
		Sensing receive ULA size & \(M=8\) \\
		Wavelength & \(\lambda=10\) mm \\
		Number of sensing snapshots & \(T=128\) \\
		Direct link separation & \(z_{\rm d}=0.05\) m \\
		Minimum MA spacing & \(d_{\min}=0.5\lambda\) \\
		Sensing noise power & \(\sigma_{\rm s}^2=-80\) dBm \\
		CU noise power & \(\sigma_k^2=-90\) dBm \\
		Target angle & \(\theta_{\rm t}=60^\circ\) \\
		CU distance & \(d_k=8\) m \\
		Target distance & \(d_{\rm t}=10\) m \\
		Target RCS & \(\sigma_{\rm RCS}=0.4\) m$^2$ \\
		AO stopping tolerance & \(\epsilon_{\rm AO}=10^{-4}\) \\
		APV inner stopping tolerance & \(\epsilon_{\rm APV}=10^{-4}\) \\
		\hline
	\end{tabular}
\end{table}
The default simulation parameters are given in Table~\ref{tab:simulation_parameters}. Without loss of generality, the pilot length is set to \(T_{\rm d}=N\), and the per-MA
pilot power is \(P_{\rm d}=2.75\) mW.  At this power, the coherent
target-reflected pilot component omitted from
\eqref{eq:pilot_received_matrix} is at least \(20\) dB below the sensing-noise
power.  For \(N=8\), the aggregate pilot power is
\(NP_{\rm d}=22\) mW, which is about \(0.7\%\) of the \(3\) W payload power
budget. 
For link distances measured in meters, we use
\({\rm PL}(d)=32.6+36.7\log_{10}(d)\) dB. Unless otherwise specified, 
the communication gains and AoDs are independently generated as
\(\alpha_{k,\ell}\sim
\mathcal{CN}(0,10^{-0.1{\rm PL}(d_k)}/L_k)\) and
\(\theta_{k,\ell}\sim\mathcal U[0,\pi]\), \(\ell=1,\ldots,L_k\).
The target coefficient is real and nonnegative, with
\(|\alpha_{\rm t}|^2
=
\sigma_{\rm RCS}
\left(10^{-0.1{\rm PL}(d_{\rm t})}\right)^2\). Furthermore, considering the sparsity of millimeter-wave channels, we set $L_k=3$, $ \forall k$ unless otherwise specified.
The evaluation set contains 1000 independently and uniformly 
drawn admissible APV errors together with all \(2^N\) vertices of
\(\mathcal E_{\rm p}(\varepsilon)\).  The
sampled worst-case CRB and the minimum SINR are evaluated over this  same evaluation set. 
We also consider three other comparable schemes: The FPA scheme fixes the transmit antennas at $\frac{\lambda}{2}$ spacing and optimizes only the beamformers without position uncertainty. The non-robust scheme jointly optimizes the beamformers and APV under \(\Delta\mathbf p=\mathbf0\). While in the oracle scheme, the position error is known before transmission. BS adjusts the commanded APV accordingly, so that the realized APV coincides with the optimized known-position geometry.

\begin{figure}[t]
\centering
\includegraphics[width=\simfigurewidth]{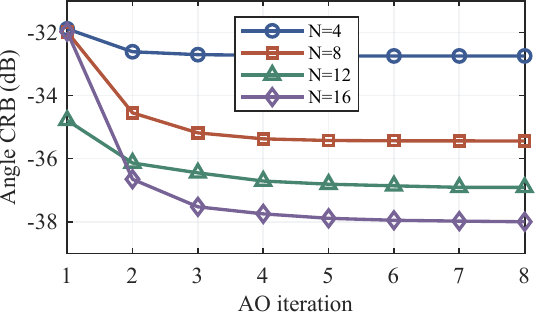}
\caption{Convergence of the proposed AO algorithm for \(K=3\).}
\label{fig:exp1_ao_convergence}
\end{figure}

Fig.~\ref{fig:exp1_ao_convergence} investigates the convergence of the
proposed AO algorithm as the number of transmit MAs changes.  We use \(K=3\)
and \(N\in\{4,8,12,16\}\) for a same representative channel realization.  All CRB sequences decrease
monotonically and become stationary within about 5 AO iterations. At convergence, the certified lower bound on CRB for
\(N=4,8,12,\) and \(16\) are \(-32.7\), \(-35.4\), \(-37.0\), and
\(-38.0\) dB, respectively. Thus, each addition of four MAs reduces the CRB
by approximately \(2.7\), \(1.6\), and \(1.2\) dB, respectively. 
These gains
confirm the benefit of the additional spatial degrees of freedom and 
transmit aperture, while their gradual reduction indicates diminishing
returns as \(N\) increases.

\begin{figure}[t]
\centering
\includegraphics[width=\simfigurewidth]{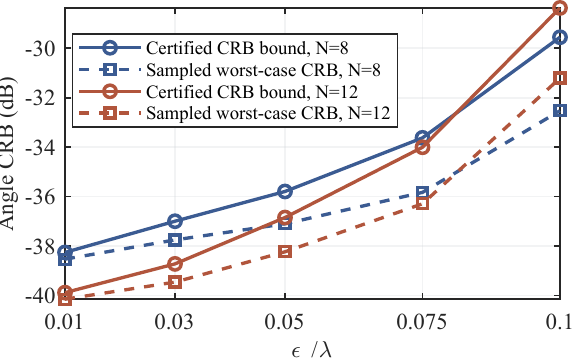}
\caption{Certified and sampled worst-case angle CRBs versus the normalized
position-error bound for \(K=3\).}
\label{fig:exp2_error_sweep}
\end{figure}
In Fig.~\ref{fig:exp2_error_sweep}, we investigate the CRB sensitivity to the MA
position error bound $\varepsilon$ for \(K=3\) and \(N\in\{8,12\}\).  We sweep
\(\varepsilon/\lambda\in\{0.01,0.03,0.05,0.075,0.10\}\) over the
independent channel realizations.  As the radius $\varepsilon/\lambda$ increases,
the certified CRB
changes from \(-38.3\) to \(-29.6\) dB for \(N=8\), and from \(-39.9\) to
\(-28.4\) dB for \(N=12\).  A larger error box weakens the worst-case pilot
information, reducing the retained
transmit-aperture information.  Additional antennas help for small and moderate
errors, but this advantage vanishes at \(0.1\lambda\) owing to the accumulated
position uncertainties.  The sampled worst-case CRB
always remains below the certified CRB bound, confirming the safe  formulation in Section \ref{subsec:rb_sensing_constraint_reformulation}.

\begin{figure}[t]
\centering
\includegraphics[width=\simfigurewidth]{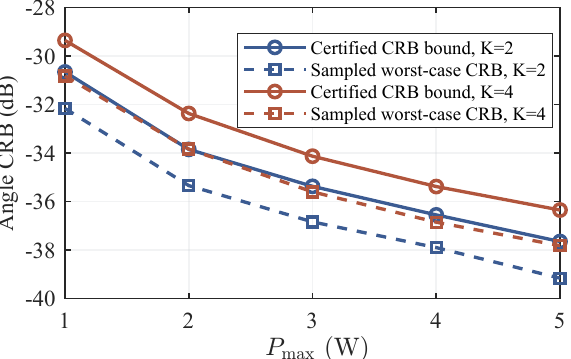}
\caption{Certified and sampled worst-case angle CRBs versus the transmit-power
budget for \(N=8\).}
\label{fig:exp3_power_sweep}
\end{figure}
Fig.~\ref{fig:exp3_power_sweep} demonstrates the CRB performance under different transmit power for \(N=8\) and \(K\in\{2,4\}\).  We set \(T_{\rm d}=N\) and vary
\(P_{\max}\in\{1,2,3,4,5\}\) W.  Increasing the budget from 1 to 5 W improves
the certified CRB by about 7 dB in both cases because the extra power increases
the target-echo energy after satisfying the communication constraints.  The \(K=2\) curve is
generally lower because fewer interference and SINR constraints leave more
beamforming freedom for sensing. 
% The nearly constant gap between the certified
%and sampled curves also shows that the safe bound does not become progressively
%looser with power.

\begin{figure}[t]
\centering
\includegraphics[width=\simfigurewidth]{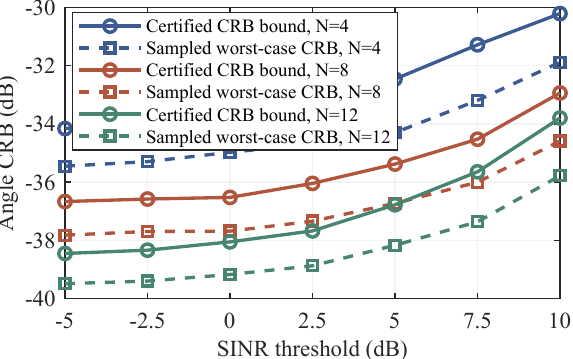}
\caption{Communication--sensing tradeoff for \(K=3\).}
\label{fig:exp4_sinr_tradeoff}
\end{figure}
Fig.~\ref{fig:exp4_sinr_tradeoff} characterizes the communication--sensing
tradeoff for \(K=3\), \(N\in\{4,8,12\}\), and a common SINR threshold from
\(-5\) to \(10\) dB.  The CRB rises
with the threshold, reaching \(-30.2\), \(-32.9\), and \(-33.8\) dB at
10 dB for \(N=4\), 8, and 12, respectively.  A tighter target redirects power
and spatial degrees of freedom from target illumination to interference
control, with the penalty becoming most pronounced at the largest thresholds.
Increasing \(N\) mitigates this loss by retaining more sensing-
aperture gain while serving the three users, particularly at the more
stringent thresholds.

\begin{figure}[t]
\centering
\includegraphics[width=\simfigurewidth]{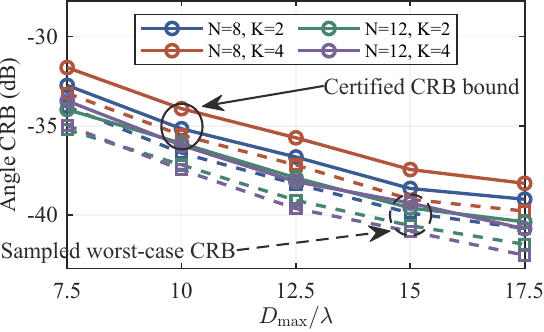}
\caption{Certified and sampled worst-case angle CRBs versus the normalized
movable-region size.}
\label{fig:exp6_movable_region_sweep}
\end{figure}
Fig.~\ref{fig:exp6_movable_region_sweep} investigates the effect of MA moving-region
size for \(N\in\{8,12\}\), \(K\in\{2,4\}\) where we vary \(D_{\max}\) over
\(\{7.5\lambda,10\lambda,12.5\lambda,15\lambda,17.5\lambda\}\).  Across all
four cases, enlarging the MA movable region improves the sampled worst-case CRB by about
6--7 dB, with \(N=12,K=4\) reaching \(-42.3\) dB. This is because the wider movable region enables
larger effective apertures and more favorable communication-channel phases.
The gain is consistent across both antenna counts and user loads.
The slight rebound for \(N=8,K=2\) at the last point reflects finite-sample
averaging and convergence to different stationary solutions; the dominant
aperture-gain trend remains clear.

\begin{figure}[t]
\centering
\includegraphics[width=\simfigurewidth]{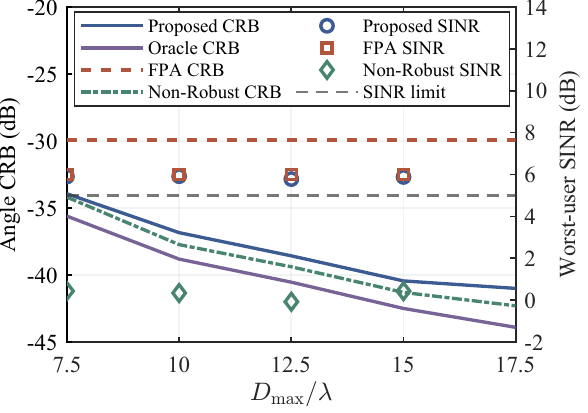}
\caption{Sensing and communication comparison.
The left and right axes show the angle CRB and sampled worst-user SINR,
respectively.}
\label{fig:exp6_scheme_comparison}
\end{figure}
Fig.~\ref{fig:exp6_scheme_comparison} compares the proposed design with other different schemes for
\(N=8\) and \(K=2\).  The proposed and non-robust designs are evaluated under
the same realized errors, whereas the FPA result is independent of
\(D_{\max}\).  As the moving region grows, the proposed design improves
its sampled worst-case CRB while maintaining a worst-user SINR of 5.8--6.1 dB
and zero sampled outage; the FPA remains at about \(-29.9\) dB and 6.0 dB.
Although the non-robust design occasionally achieves a lower CRB, its
worst-user SINR is only \(-0.1\) to 0.5 dB and its outage probability is
nearly one half, violating the 5-dB requirement.  Its apparent sensing gain
therefore comes at the cost of communication reliability, while the oracle
achieves the lowest CRB by removing position uncertainty before optimization.

% \begin{figure}[t]
% \centering
% \includegraphics[width=\simfigurewidth]{exp7_estimator_validation.pdf}
% \caption{Angle-estimation MSE and the corresponding CRBs versus radar SNR.}
% \label{fig:exp7_estimator_validation}
% \end{figure}
% Fig.~\ref{fig:exp7_estimator_validation} validates the exact joint CRB against
% the angle mean-squared error (MSE) of the joint ML estimator.  We fix the
% \(N=8\), \(K=3\), \(\varepsilon=0.05\lambda\) design obtained for the
% first channel realization, sweep the target-echo signal-to-noise ratio (SNR)
% from \(-10\) to 20 dB in 5-dB increments, and perform 200 position-error and
% noise trials per point.  The joint estimator is compared with the oracle-APV,
% nominal-APV, and receive-only estimators over the angular interval
% \([5^\circ,175^\circ]\).  Throughout the SNR range, the joint-ML MSE closely
% tracks its CRB and nearly coincides with the oracle-APV MSE, reaching about
% \(-83\) dB rad\(^2\) at 20 dB.
% By contrast, the nominal-APV estimator exhibits an error floor near \(-56\) dB
% rad\(^2\) owing to position-model mismatch.  The receive-only estimator keeps
% improving with SNR but is about 17 dB worse than joint ML at 20 dB because it
% discards the transmit-aperture information.  These results verify the CRB and
% the benefit of pilot-assisted joint estimation.

\begin{figure}[t]
\centering
\includegraphics[width=\simfigurewidth]{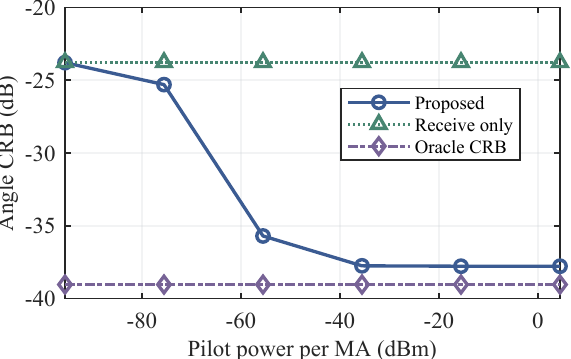}
\caption{Worst-error angle CRB versus the pilot power.}
\label{fig:exp8_pilot_recovery}
\end{figure}
Fig.~\ref{fig:exp8_pilot_recovery} illustrates how pilot power recovers usable
transmit-aperture information.  We consider a single deterministic single-path
scene with \(N=8\), \(K=2\) and equal CU path gains. The CU angles are set as \(100^\circ\) and
\(165^\circ\), and \(\varepsilon=0.05\lambda\).  The proposed design is
fixed for this scene while the 2.75-mW pilot power is scaled by factors in
\(\{10^{-10},10^{-8},10^{-6},10^{-4},10^{-2},1\}\).  Its CRB is evaluated over
the position-error set together with the receive-only limit and the
independently optimized oracle bound.  The proposed CRB improves
from \(-23.8\) to \(-37.8\) dB, moving from the receive-only limit of
\(-23.8\) dB toward the oracle bound of \(-39.0\) dB.  Meanwhile, the mean
retained-aperture fraction grows from \(1.6\times10^{-4}\) to nearly one.  Since the
position Fisher information provided by the pilot scales with its power, weak
pilots provide little information for separating angular phase from position
errors, whereas sufficiently informative pilots recover almost the entire transmit
aperture, confirming the aperture-recovery mechanism predicted by the CRB.

\begin{figure}[t]
\centering
\makebox[\simfigurewidth][c]{%
\subfloat[]{%
\includegraphics[height=3.9cm]{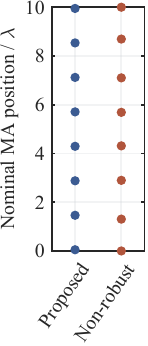}%
\label{fig:exp11_apv_positions}}
\hfill\hspace{0.3cm}
\subfloat[]{%
\includegraphics[height=3.9cm]{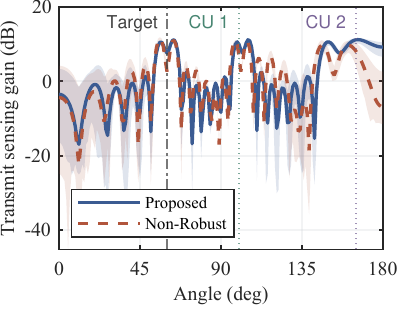}%
\label{fig:exp11_sensing_response}}}
\caption{(a) Designed nominal MA positions and (b) Beam pattern of the proposed and non-robust designs.}
\label{fig:exp11_apv_response}
\end{figure}
Fig.~\ref{fig:exp11_apv_response} visualizes the optimized MA geometry and
angular response for the same deterministic single-path scene considered in
Fig.~\ref{fig:exp8_pilot_recovery}.  The curves show nominal responses and
the shaded regions span the minimum-to-maximum gains over all evaluated
position errors.  The
proposed APV is nearly uniformly spread while retaining robust boundary
margins.  Its nominal and worst-error target gains are 7.6 and 7.3 dB,
compared with 2.6 and 2.5 dB for the non-robust design.  Its lower-envelope
gain is also higher at both CU angles, particularly at \(165^\circ\), where the
two designs achieve 10.2 and 4.1 dB, respectively.  Joint geometry and
beamforming optimization thus stabilizes the desired responses, whereas the
non-robust design relies on nominal constructive phases that may be lost after
displacement.

\begin{figure}[t]
\centering
\subfloat[]{%
\includegraphics[width=6.1cm]{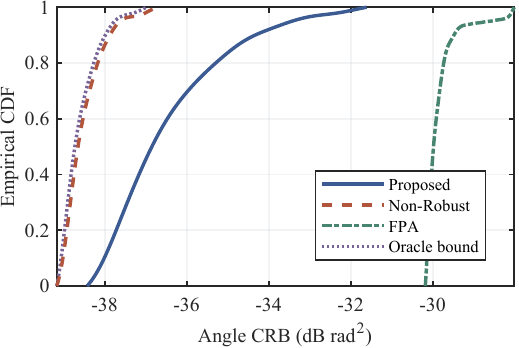}%
\label{fig:exp12_sensing_cdf}}\\
\subfloat[]{%
\includegraphics[width=6.1cm]{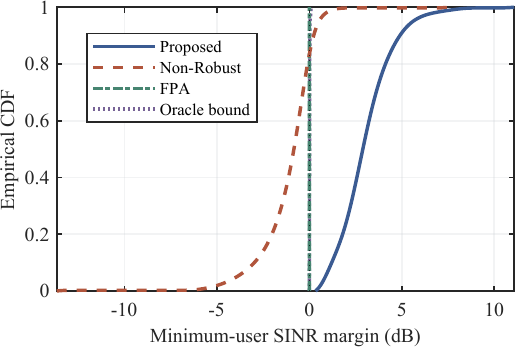}%
\label{fig:exp12_sinr_cdf}}
\caption{Smoothed empirical CDFs of (a) the angle CRB and (b) the
minimum-user SINR margin under random position errors.}
\label{fig:exp12_reliability_cdf}
\end{figure}
Fig.~\ref{fig:exp12_reliability_cdf} evaluates sensing and communication
reliability using smoothed empirical cumulative distribution functions (CDFs).  For 200
independent channel scenes with \(N=8\), \(K=3\) and a 5-dB SINR
requirement, 2000 errors are drawn uniformly from the error box per scene, yielding
400,000 samples for each scheme.  All schemes use the same channel scenes, and
the proposed and non-robust designs also use the same realized errors.  The
median CRBs of the proposed, non-robust,
FPA, and oracle designs are \(-36.9\), \(-38.7\), \(-30.0\), and
\(-38.7\) dB rad\(^2\), respectively.  Despite its lower CRB, the non-robust
design has an 83.6\% outage probability, a \(-0.8\)-dB median SINR margin,
and a \(-3.9\)-dB 5th-percentile margin.  The proposed design instead achieves
zero outage, a 2.9-dB median margin, and a 0.3-dB worst observed margin.  The
FPA and oracle also meet the SINR target but operate close to the active
constraint.  Thus, the proposed method trades part of the ideal sensing gain
for reliable communication, whereas the oracle assumes unavailable error
knowledge and the FPA sacrifices movable-aperture gain.

\section{Conclusion}
\label{sec:conclusion}
This paper studied robust beamforming and nominal APV design for a multiuser
MA-ISAC system with unknown MA position errors. We introduced a
pilot-assisted protocol that uses the low-power near-field leakage and echoes to  support the joint estimation of the target angle, reflection
coefficient, and APV error.
% The resulting angle CRB characterizes how
%noisy position information from the pilot preserves the movable
%transmit-aperture contribution, while the fixed receive aperture remains
%directly usable.
 We formulated a worst-case CRB minimization problem subject to
robust multiuser SINR requirements and practical deployment constraints. A common phase-response decomposition and analytical uncertainty
bounds led to an alternating algorithm with beam and APV updates.
% Numerical results verified the CRB and SINR guarantees and demonstrated
% improved communication reliability over nominal-position designs.
Numerical results demonstrated lower angle CRBs and improved communication
reliability over nominal-position designs while satisfying the prescribed SINR
requirements. Future extension of this work
may extend the proposed framework to wideband calibration and  multiple targets sensing scenarios.
\appendices
\section{Proof of Proposition~\ref{prop:rb_exact_aperture_decomposition}}
\label{app:crb_derivation}

Write \(\alpha_{\rm t}=\alpha_{\rm R}+j\alpha_{\rm I}\) and collect the real
unknowns as
\(\boldsymbol\xi\triangleq
[\theta_{\rm t},\alpha_{\rm R},\alpha_{\rm I},
\Delta\mathbf p^{\rm T}]^{\rm T}\).
Let \(\overline{\mathbf Y}_{\rm d}\) and
\(\overline{\mathbf Y}_{\rm s}\) be the noise-free means in
\eqref{eq:pilot_received_matrix} and
\eqref{eq:sensing_received_matrix}.  After normalizing the two observations by the noise standard deviation, define the Jacobian
\begin{equation}
\mathbf D\triangleq
\frac{1}{\sigma_{\rm s}}
\frac{\partial}{\partial\boldsymbol\xi^{\rm T}}
\begin{bmatrix}
\operatorname{vec}(\overline{\mathbf Y}_{\rm d})\\
\operatorname{vec}(\overline{\mathbf Y}_{\rm s})
\end{bmatrix}.
\end{equation}
The corresponding Fisher information matrix (FIM) is
\(\mathbf J_{\boldsymbol\xi}=2\Re\{\mathbf D^{\rm H}\mathbf D\}\).
The pilot mean depends only on the position errors, with
$\frac{\partial\overline{\mathbf Y}_{\rm d}}{\partial\Delta p_n}
=
\frac{\partial\mathbf h_{\rm d}(\tilde p_n)}{\partial\tilde p_n}
[\mathbf X_{\rm d}]_{n,:}$.
Using the pilot orthogonality in \eqref{eq:pilot_covariance} directly gives \eqref{eq:pilot_position_information}.
For the payload observation, define
\(\dot{\mathbf b}\triangleq
\left.\partial\mathbf b(\theta)/\partial\theta\right|_{\theta=\theta_{\rm t}}\).
The two reflection-coefficient derivatives are
\(\mathbf b\mathbf u^{\rm H}\mathbf S\) and
\(j\mathbf b\mathbf u^{\rm H}\mathbf S\).  The angle and position derivatives are
\begin{subequations}
\label{eq:app_joint_mean_derivatives}
\begin{align}
\frac{\partial\overline{\mathbf Y}_{\rm s}}{\partial\theta_{\rm t}}
&=\alpha_{\rm t}
\left(
\dot{\mathbf b}\mathbf u^{\rm H}
+\mathbf b\mathbf v^{\rm H}
\right)\mathbf S,
\label{eq:app_sensing_angle_derivative}\\
\frac{\partial\overline{\mathbf Y}_{\rm s}}{\partial\Delta p_n}
&=\alpha_{\rm t}\mathbf b
(\mathbf G\mathbf e_n)^{\rm H}\mathbf S.
\label{eq:app_sensing_position_derivative}
\end{align}
\end{subequations}
Let \(\mathbf P_{\mathbf b}^{\perp}\triangleq
\mathbf I_M-\mathbf b\mathbf b^{\rm H}/M\).  The receive component
\(\mathbf P_{\mathbf b}^{\perp}\dot{\mathbf b}\) is orthogonal to both
reflection-coefficient derivatives and satisfies
\(\|\mathbf P_{\mathbf b}^{\perp}\dot{\mathbf b}\|_2^2=\eta_{\rm r}\).
Eliminating the reflection coefficient while retaining
\((\theta_{\rm t},\Delta\mathbf p)\) projects the transmit-side angle and
position derivatives onto \(\mathbf P_{\mathbf u}^{\perp}\).  Specifically,
the projected payload derivatives are
\begin{subequations}
\label{eq:app_projected_payload_derivatives}
\begin{align}
\mathbf D_{\theta}^{\perp}
&=\alpha_{\rm t}\left[
\mathbf P_{\mathbf b}^{\perp}\dot{\mathbf b}\mathbf u^{\rm H}
+\mathbf b(\mathbf P_{\mathbf u}^{\perp}\mathbf v)^{\rm H}
\right]\mathbf S,\\
\mathbf D_{p_n}^{\perp}
&=\alpha_{\rm t}\mathbf b
(\mathbf P_{\mathbf u}^{\perp}\mathbf G\mathbf e_n)^{\rm H}\mathbf S.
\end{align}
\end{subequations}
The two components of \(\mathbf D_{\theta}^{\perp}\) are orthogonal because
\(\mathbf P_{\mathbf b}^{\perp}\dot{\mathbf b}\perp\mathbf b\).  Hence,
eliminating \((\alpha_{\rm R},\alpha_{\rm I})\) and adding the pilot
contribution yields the following equivalent FIM for
\((\theta_{\rm t},\Delta\mathbf p)\):
\begin{equation}
\left[
\begin{array}{cc}
\gamma_{\rm s}\eta_{\rm r}\|\mathbf u\|_2^2
+\gamma_{\rm s}M
\|\mathbf P_{\mathbf u}^{\perp}\mathbf v\|_2^2
&
\gamma_{\rm s}M\Re\!\left\{
\mathbf v^{\rm H}\mathbf P_{\mathbf u}^{\perp}\mathbf G
\right\}
\\
\gamma_{\rm s}M\Re\!\left\{
\mathbf G^{\rm H}\mathbf P_{\mathbf u}^{\perp}\mathbf v
\right\}
&
\mathbf J_{\rm d}+\mathbf J_{\rm e}
\end{array}
\right].
\label{eq:app_reduced_angle_position_fim}
\end{equation}
Using \(\mathbf v=-\tan\theta_{\rm t}\mathbf G\tilde{\mathbf p}\) and
\eqref{eq:echo_position_information}, the transmit-angle term and the
angle-position cross term reduce to
\(\tan^2\theta_{\rm t}\tilde{\mathbf p}^{\rm T}\mathbf J_{\rm e}
\tilde{\mathbf p}\) and
\(-\tan\theta_{\rm t}\mathbf J_{\rm e}\tilde{\mathbf p}\), respectively.
Eliminating \(\Delta\mathbf p\) by taking the Schur complement of the
lower-right block in \eqref{eq:app_reduced_angle_position_fim} and using
\(\mathbf J_{\rm e}-\mathbf J_{\rm e}
(\mathbf J_{\rm e}+\mathbf J_{\rm d})^{-1}\mathbf J_{\rm e}
=\mathbf J_{\rm e}(\mathbf J_{\rm e}+\mathbf J_{\rm d})^{-1}\mathbf J_{\rm d}\)
yields
\begin{equation}
\relax[\mathbf J_{\boldsymbol\xi}^{-1}]_{1,1}^{-1}
=\gamma_{\rm s}\eta_{\rm r}\|\mathbf u\|_2^2
+\tan^2\theta_{\rm t}\,
\tilde{\mathbf p}^{\rm T}\mathbf J_{\rm e}
(\mathbf J_{\rm e}+\mathbf J_{\rm d})^{-1}
\mathbf J_{\rm d}\tilde{\mathbf p}.
\label{eq:app_exact_crb_reduction}
\end{equation}
Finally, Lagrange's identity gives
\begin{equation}
\begin{aligned}
\|\mathbf u\|_2^2
\|\mathbf P_{\mathbf u}^{\perp}\mathbf v\|_2^2
&=\|\mathbf u\|_2^2\|\mathbf v\|_2^2
-|\mathbf u^{\rm H}\mathbf v|^2\\
&=\sum_{k<\ell}|v_k u_\ell-v_\ell u_k|^2
=\|\mathbf d\|_2^2.
\end{aligned}
\label{eq:app_lagrange_identity}
\end{equation}
Combining this identity with
\(\mathbf v=-\tan\theta_{\rm t}\mathbf G\tilde{\mathbf p}\) and
\eqref{eq:echo_position_information} rewrites the second term in
\eqref{eq:app_exact_crb_reduction} in the transmit-aperture form of
\eqref{eq:crb_theta}.  Taking the reciprocal completes the proof.

\section{Proof of Lemma~\ref{lem:rb_tight_phase_remainder}}
\label{app:proof_tight_phase_remainder}

Direct expansion gives
$|e^{jz}-c(x)-jb(x)z|^2-a^2(x)=(\sin z-b(x)z)^2
-(1-\cos z)(\cos z-\cos x)$.
For \(0<x<\pi\), let
\(F_x(z)=a^2(x)-|e^{jz}-c(x)-jb(x)z|^2\).  For \(z>0\),
\begin{equation}
\frac{F_x'(z)}{2z}
=(b(x)-c(x))\frac{\sin z}{z}+b(x)\cos z-b^2(x).
\end{equation}
Because \(b(x)/c(x)=2\tan(x/2)/x\ge1\), while \(\sin z/z\) and \(\cos z\)
decrease on \((0,x]\), the right-hand side decreases strictly from
\(2b(x)-c(x)-b^2(x)>0\) to \(-a(x)b(x)<0\).  Hence, \(F_x\)
first increases and then decreases; together with \(F_x(0)=F_x(x)=0\) and
evenness, this proves \(F_x(z)\ge0\) for \(|z|\le x\).
%This proves \eqref{eq:rb_affine_phase_response}.

\section{Proof of Proposition~\ref{prop:rb_positive_response_lower_bound}}
\label{app:proof_positive_response_lower_bound}

Consider the generic phase-weighted response
\begin{equation}
f(\Delta\mathbf p)
\triangleq
(\alpha+\boldsymbol\beta^{\rm T}\Delta\mathbf p)
e^{j\boldsymbol\omega^{\rm T}\Delta\mathbf p},
\end{equation}
and define
\(\varphi\triangleq
\varepsilon\|\boldsymbol\omega\|_1<\pi\).
For every \(\Delta\mathbf p\in\mathcal E_{\rm p}(\varepsilon)\),
\(|\boldsymbol\omega^{\rm T}\Delta\mathbf p|\le\varphi\).
Applying Lemma~\ref{lem:rb_tight_phase_remainder} with
\(z=\boldsymbol\omega^{\rm T}\Delta\mathbf p\) to the exponential factor
in \(f(\Delta\mathbf p)\) directly yields
\begin{equation}
f(\Delta\mathbf p)
=c(\varphi)(\alpha+\boldsymbol\beta^{\rm T}\Delta\mathbf p)
+jb(\varphi)\alpha\boldsymbol\omega^{\rm T}\Delta\mathbf p+\epsilon,
\label{eq:app_affine_amplitude_phase_decomposition}
\end{equation}
where
\(|r_{\varphi}(\boldsymbol\omega^{\rm T}\Delta\mathbf p)|
\le a(\varphi)\), and the residual term satisfies
$\epsilon=
(\alpha+\boldsymbol\beta^{\rm T}\Delta\mathbf p)
r_{\varphi}(\boldsymbol\omega^{\rm T}\Delta\mathbf p)+jb(\varphi)(\boldsymbol\beta^{\rm T}\Delta\mathbf p)
(\boldsymbol\omega^{\rm T}\Delta\mathbf p)$,
where $|\epsilon|\le{}
\bigl(|\alpha|+|\boldsymbol\beta^{\rm T}\Delta\mathbf p|\bigr)
a(\varphi)
+b(\varphi)|\boldsymbol\beta^{\rm T}\Delta\mathbf p|
|\boldsymbol\omega^{\rm T}\Delta\mathbf p|$.
Applying \eqref{eq:app_affine_amplitude_phase_decomposition} to  \(\mathbf q^{\rm H}\mathbf u\) with \(\alpha=1\),
\(\boldsymbol\beta=\mathbf0\), and
\(\boldsymbol\omega=\kappa_{\rm t}\mathbf e_n\), yields
\begin{equation}
\begin{aligned}
\Re\{\mathbf q^{\rm H}\mathbf u\}
\ge{}&c(\varphi_{\rm t})\Re\{\mathbf1_N^{\rm T}\mathbf z_u\}
-b(\varphi_{\rm t})\kappa_{\rm t}
\Im\{\mathbf z_u\}^{\rm T}\Delta\mathbf p\\
&-a(\varphi_{\rm t})\|\mathbf z_u\|_1.
\end{aligned}
\label{eq:app_receive_response_lower_bound}
\end{equation}
Next, expanding \(d_{k,\ell}=v_ku_\ell-v_\ell u_k\) over the transmit
antennas and collecting the antenna pairs gives the exact response
\begin{equation}
\mathbf d(\mathbf p+\Delta\mathbf p)
=\mathbf C_d\!\left[
\mathbf B_{\Delta}(\mathbf p+\Delta\mathbf p)
\odot
e^{j\kappa_{\rm t}\mathbf B_{\Sigma}\Delta\mathbf p}
\right],
\label{eq:app_exact_pair_response}
\end{equation}
where the exponential is evaluated entrywise.  Applying
\eqref{eq:app_affine_amplitude_phase_decomposition} to its entries with
\(\alpha=p_n-p_m\),
\(\boldsymbol\beta=\mathbf e_n-\mathbf e_m\), and
\(\boldsymbol\omega=\kappa_{\rm t}(\mathbf e_n+\mathbf e_m)\) yields
\begin{equation}
\mathbf d(\mathbf p+\Delta\mathbf p)
=\bar{\mathbf d}+\mathbf A_{\rm d}\Delta\mathbf p
+\mathbf C_d\boldsymbol\epsilon_d,
\,\,\,\,
|\boldsymbol\epsilon_d|\preceq\boldsymbol\rho_d,
\label{eq:rb_pair_response_decomposition}
\end{equation}
where \(\bar{\mathbf d}\), \(\mathbf A_{\rm d}\), and
\(\boldsymbol\rho_d\) are defined in
Proposition~\ref{prop:rb_positive_response_lower_bound}.  For this
specialization, \(\varphi=2\varphi_{\rm t}\),
\(|\alpha|=p_m-p_n\), and
\(|\boldsymbol\beta^{\rm T}\Delta\mathbf p|\le 2\varepsilon\).
Moreover, the remaining product satisfies
$|(\Delta p_n-\Delta p_m)(\Delta p_n+\Delta p_m)|
=|\Delta p_n^2-\Delta p_m^2|
\le\varepsilon^2$.
These relations give the entries of \(\boldsymbol\rho_d\).
Consequently,
\begin{equation}
\Re\{\mathbf t^{\rm H}\mathbf d\}
\ge \Re\{\mathbf t^{\rm H}\bar{\mathbf d}\}
+\Re\{\mathbf A_{\rm d}^{\rm H}\mathbf t\}^{\rm T}\Delta\mathbf p
-\boldsymbol\rho_d^{\rm T}|\mathbf C_d^{\rm H}\mathbf t|.
\label{eq:app_transmit_response_lower_bound}
\end{equation}
 Using
\(\min_{\Delta\mathbf p\in\mathcal E_{\rm p}(\varepsilon)}
\mathbf g^{\rm T}\Delta\mathbf p=-\varepsilon\|\mathbf g\|_1\) and
combining \eqref{eq:app_receive_response_lower_bound} with
\eqref{eq:app_transmit_response_lower_bound} and completes the proof.

\section{APV Differentials and Quadratic Lower Models}
\label{app:analytic_apv_models}

This appendix provides the two ingredients required by
\eqref{eq:rb_paired_apv_models}: the APV-dependent differentials and a unified
curvature matrix that guarantees a valid quadratic lower model. In the following, let
\(\mathrm d\) denote the differential.

\subsection{APV-Dependent Differentials}

For
\(r_m^{\rm d}(x)\triangleq
\sqrt{(x-(m-1)\lambda/2)^2+z_{\rm d}^2}\), differentiating
\eqref{eq:near_field_direct_channel} gives
$	g_{\rm d}(x)
	=\frac{2T_{\rm d}P_{\rm d}}{\sigma_{\rm s}^2}
	\left(\frac{\lambda}{4\pi}\right)^2
	\sum_{m=1}^{M}
	\frac{\left(x-\frac{(m-1)\lambda}{2}\right)^2}{[r_m^{\rm d}(x)]^4}\times
	\left[\kappa^2+\frac{1}{[r_m^{\rm d}(x)]^2}\right]$.
The differential of \(\underline g_{\rm d}(p_n)\) follows from its active
minimizing branch: it is
\(g_{\rm d}'(p_n\pm\varepsilon)\mathrm d p_n\) at an active endpoint
and zero at an interior minimizer.  At a tie, either active subderivative is
valid, while the larger adjacent curvature bound is used; branch changes are
handled by the APV backtracking step.  These derivatives populate only the
lower-right block of \(\mathrm d\mathbf M_{\rm s}\).

%For the lower bound \(\underline\Phi(\mathbf W,\mathbf p)\) in \eqref{eq:rb_phi}, \(\mathrm d\mathbf z_u\) and \(\mathrm d\mathbf C_d\) are given by
%\begin{subequations}
%\begin{equation}
%	\begin{aligned}
%		\mathrm d\mathbf z_u
%		&=j\kappa_{\rm t}\operatorname{diag}(\mathbf z_u)\mathrm d\mathbf p,\\
%		\mathrm d\mathbf C_d
%		&=j\kappa_{\rm t}\mathbf C_d
%		\operatorname{diag}(\mathbf B_{\Sigma}\mathrm d\mathbf p).
%	\end{aligned}
%\end{equation}
%\end{subequations}
%The differentials \(\mathrm d\bar{\mathbf d}\) and
%\(\mathrm d\mathbf A_{\rm d}\) follow by differentiating
%\eqref{eq:rb_pair_affine_center} and
%\eqref{eq:rb_pair_affine_jacobian}, respectively. Since \(p_n<p_m\) for every
%\(1\le n<m\le N\),
%\begin{equation}
%\mathrm d\boldsymbol\rho_d
%=-a(2\varphi_{\rm t})
%\mathbf B_{\Delta}\mathrm d\mathbf p,
%\,\,\,\,
%\mathrm d^2\boldsymbol\rho_d=\mathbf0.
%\end{equation}
%The magnitudes of \(\mathbf z_u\) and
%\(\mathbf C_d^{\rm H}\mathbf t\) are constant with
%respect to \(\mathbf p\).
%On a smooth branch, the differential of \(\underline\Phi\) in
%\eqref{eq:rb_phi} follows from the preceding identities and standard
%differentiation rules.

For  \(\underline\Phi(\mathbf W,\mathbf p)\) in \eqref{eq:rb_phi}, the related differentials are
\begin{subequations}
	\label{eq:app_phi_basic_differentials}
	\begin{align}
		\mathrm d\mathbf z_u
		&=
		j\kappa_{\rm t}
		\operatorname{diag}(\mathbf z_u)\mathrm d\mathbf p,
		\\
		\mathrm d\mathbf C_d
		&=
		j\kappa_{\rm t}\mathbf C_d
		\operatorname{diag}(\mathbf B_{\Sigma}\mathrm d\mathbf p),
		\\
		\mathrm d\bar{\mathbf d}
		&=
		c(2\varphi_{\rm t})
		\left(
		\mathrm d\mathbf C_d\mathbf B_{\Delta}\mathbf p
		+\mathbf C_d\mathbf B_{\Delta}\mathrm d\mathbf p
		\right),
		\\
		\mathrm d\mathbf A_{\rm d}
		&=
		\mathrm d\mathbf C_d
		\left[
		c(2\varphi_{\rm t})\mathbf B_{\Delta}
		+jb(2\varphi_{\rm t})\kappa_{\rm t}
		\operatorname{diag}(\mathbf B_{\Delta}\mathbf p)
		\mathbf B_{\Sigma}
		\right]
		\notag\\
		&\quad+
		jb(2\varphi_{\rm t})\kappa_{\rm t}\mathbf C_d
		\operatorname{diag}(\mathbf B_{\Delta}\mathrm d\mathbf p)
		\mathbf B_{\Sigma},
		\\
		\mathrm d\boldsymbol\rho_d
		&=
		-a(2\varphi_{\rm t})
		\mathbf B_{\Delta}\mathrm d\mathbf p,
		\qquad
		\mathrm d^2\boldsymbol\rho_d=\mathbf0,
		\\
		\mathrm d\mathbf g
		&=
		M\Re\!\left\{
		(\mathrm d\mathbf A_{\rm d})^{\rm H}\mathbf t
		\right\}
		-\eta_{\rm r}b(\varphi_{\rm t})\kappa_{\rm t}
		\Im\{\mathrm d\mathbf z_u\}.
	\end{align}
\end{subequations}
Define
\(\mathbf s_g\triangleq\operatorname{sign}(\mathbf g)\) on a smooth branch of
\(\|\mathbf g\|_1\).
The differential of \(\underline\Phi\) is then
\begin{equation}
	\begin{aligned}
		\mathrm d\underline\Phi
		=2\Big[
		&\eta_{\rm r}c(\varphi_{\rm t})
		\Re\!\left\{
		\mathbf1_N^{\rm T}\mathrm d\mathbf z_u
		\right\}
		+M\Re\!\left\{
		\mathbf t^{\rm H}\mathrm d\bar{\mathbf d}
		\right\}
		-\varepsilon\mathbf s_g^{\rm T}\mathrm d\mathbf g
		\\
		&-M
		(\mathrm d\boldsymbol\rho_d)^{\rm T}
		|\mathbf C_d^{\rm H}\mathbf t|
		\Big].
	\end{aligned}
	\label{eq:app_phi_differential}
\end{equation}

For
\(\mathbf M_u\) in \eqref{eq:rb_echo_trust_region_lmi}, one has
\begin{equation}
\mathrm d\bar{\mathbf u}
=j\kappa_{\rm t}\cos\varphi_{\rm t}\,
\mathbf W^{\rm H}
\operatorname{diag}(\mathbf a_{\rm t}(\mathbf p))
\mathrm d\mathbf p.
\label{eq:app_echo_center_differential}
\end{equation}
With \((\overline U,\lambda_u)\) fixed,
\(\mathrm d\mathbf M_u\) is zero except in the two off-diagonal blocks
containing \(\bar{\mathbf u}\).

In \eqref{eq:rb_robust_comm_lmi}, we differentiate \(\mathbf M_k\)  through its channel center and ellipsoid factor.
With \(\boldsymbol\Omega_k\) and \(\nu_k\) fixed in the current subproblem, \eqref{eq:rb_comm_channel_center}--\eqref{eq:rb_comm_channel_jacobian}
give
\begin{subequations}
\begin{align}
\mathrm d[\bar{\mathbf h}_k]_n
&=j\!\sum\nolimits_{\ell=1}^{L_k}\alpha_{k,\ell}c(\varphi_{k,\ell})
\kappa_{k,\ell}e^{j\kappa_{k,\ell}p_n}\,\mathrm d p_n,\\
\mathrm d[\mathbf J_k]_{n,n}
&=-\!\sum\nolimits_{\ell=1}^{L_k}\alpha_{k,\ell}b(\varphi_{k,\ell})
\kappa_{k,\ell}^2e^{j\kappa_{k,\ell}p_n}\,\mathrm d p_n,\\
\mathrm d\mathbf R_k
&=\varepsilon\sqrt{\operatorname{tr}(\boldsymbol\Omega_k)}
\begin{bmatrix}
\Re\{\mathrm d\mathbf J_k(\boldsymbol\Omega_k^\dagger)^{1/2}\}\\
\Im\{\mathrm d\mathbf J_k(\boldsymbol\Omega_k^\dagger)^{1/2}\}
\end{bmatrix}.
\end{align}
\end{subequations}
To differentiate the matrix square root in \eqref{eq:rb_comm_shape_matrix}, let
\begin{equation}
\mathbf P_k\triangleq
(1+\nu_k)\mathbf R_k\mathbf R_k^{\rm T}
+(1+\nu_k^{-1})\rho_k^2\mathbf I_{2N}.
\end{equation}
Since \(\rho_k\) is independent of \(\mathbf p\),
\begin{subequations}
\begin{gather}
\mathrm d\mathbf P_k
=(1+\nu_k)\left(
\mathrm d\mathbf R_k\mathbf R_k^{\rm T}
+\mathbf R_k\mathrm d\mathbf R_k^{\rm T}\right),\\
\mathbf P_k^{1/2}\mathrm d(\mathbf P_k^{1/2})
+\mathrm d(\mathbf P_k^{1/2})\mathbf P_k^{1/2}
=\mathrm d\mathbf P_k,\\
\mathrm d\mathbf E_k
=[\mathbf I_N,j\mathbf I_N]\mathrm d(\mathbf P_k^{1/2}),\\
\mathrm d\mathbf T_k
=[\mathrm d\mathbf E_k,\mathrm d\bar{\mathbf h}_k],\\
\mathrm d\mathbf M_k
=\Re\!\left\{
(\mathrm d\mathbf T_k)^{\rm H}\mathbf S_k\mathbf T_k
+\mathbf T_k^{\rm H}\mathbf S_k\mathrm d\mathbf T_k
\right\}.
\end{gather}
\end{subequations}

\subsection{Curvature Matrix and Safe Lower-Bound}

For a scalar, vector, or matrix
function \(\mathbf X(\mathbf p)\), let \(v_{\mathbf X}\),
\(d_{\mathbf X,n}\), and \(h_{\mathbf X,n,m}\) be bounds on
\(\|\mathbf X\|_2\), \(\|\partial\mathbf X/\partial p_n\|_2\), and
\(\|\partial^2\mathbf X/(\partial p_n\partial p_m)\|_2\), respectively,
over the smooth algebraic branch containing \(\mathbf p^{(r,i)}\).  For the
generic phase-weighted affine responses appearing in the APV-dependent
quantities,
\(f(\mathbf p)=(\alpha+\boldsymbol\beta^{\rm T}\mathbf p)
e^{j\boldsymbol\omega^{\rm T}\mathbf p}\),
\(\mathbf p\in[0,D_{\max}]^N\), valid choices are
\begin{subequations}
	\label{eq:curvature_bounds}
	\begin{align}
		v_f
		&=|\alpha|+D_{\max}\|\boldsymbol\beta\|_1,
		\\
		d_{f,n}
		&=|\beta_n|+|\omega_n|v_f,
		\\
		h_{f,n,m}
		&=|\omega_n\beta_m+\omega_m\beta_n|
		+|\omega_n\omega_m|v_f.
	\end{align}
\end{subequations}
%Bounds for sums, products, scalar compositions, Hermitian transposition, and
%real-part extraction are propagated using the standard norm, product, and
%chain rules, with interval endpoint bounds used for square-root and reciprocal
%terms.
Following the standard derivative bound propagation rules
\cite{griewank2008evaluating}, bounds for sums, linear mappings and real-part
extraction follow directly from the triangle inequality.  For the
product
\(\mathbf Z(\mathbf p)=\mathbf X(\mathbf p)\mathbf Y(\mathbf p)\),
the spectral-norm bounds are propagated as
\begin{subequations}
	\label{eq:app_curvature_product_rules}
	\begin{align}
		v_{\mathbf Z}
		&\le v_{\mathbf X}v_{\mathbf Y},\\
		d_{\mathbf Z,n}
		&\le
		d_{\mathbf X,n}v_{\mathbf Y}
		+v_{\mathbf X}d_{\mathbf Y,n},\\
		h_{\mathbf Z,n,m}
		&\le
		h_{\mathbf X,n,m}v_{\mathbf Y}
		+d_{\mathbf X,n}d_{\mathbf Y,m}\notag\\
		&\quad+
		d_{\mathbf X,m}d_{\mathbf Y,n}
		+v_{\mathbf X}h_{\mathbf Y,n,m}.
	\end{align}
\end{subequations}
Apply the recursions to
\(\mathbf F\in\{\underline\Phi,\mathbf M_{\rm s},\mathbf M_u,\mathbf M_k\}\)
in \eqref{prob:rb_apv_update}.  If
\(h_{\mathbf F,n,m}\) is the resulting bound on the corresponding second
derivative, the curvature matrix in \eqref{eq:rb_paired_apv_models} is selecte as
\begin{equation}
\mathbf H_{\mathbf F}^{(r,i)}
\triangleq
\operatorname{diag}\!\left(
\sum_{m=1}^{N}h_{\mathbf F,1,m},\ldots,
\sum_{m=1}^{N}h_{\mathbf F,N,m}
\right).
\end{equation}
Let
\(\boldsymbol\delta\triangleq\mathbf p-\mathbf p^{(r,i)}\).  After symmetrizing
the bounds so that \(h_{\mathbf F,n,m}=h_{\mathbf F,m,n}\),
\begin{align}
&\left\|
\sum_{n,m}\delta_n\delta_m
\frac{\partial^2\mathbf F}{\partial p_n\partial p_m}
\right\|_2
\le
\sum_{n,m}h_{\mathbf F,n,m}|\delta_n\delta_m|\notag\\
&\hspace{18mm}\le
\sum_n\left(\sum_m h_{\mathbf F,n,m}\right)\delta_n^2
=\boldsymbol\delta^{\rm T}\mathbf H_{\mathbf F}^{(r,i)}
\boldsymbol\delta,
\end{align}
where the second inequality follows from
\(2|\delta_n\delta_m|\le\delta_n^2+\delta_m^2\).  Since the directional
second derivative is Hermitian,
\begin{equation}
\sum_{n,m}\delta_n\delta_m
\frac{\partial^2\mathbf F}{\partial p_n\partial p_m}
\succeq
-\boldsymbol\delta^{\rm T}\mathbf H_{\mathbf F}^{(r,i)}
\boldsymbol\delta\,\mathbf I_m.
\end{equation}
Integrating this inequality along the segment from \(\mathbf p^{(r,i)}\) to
\(\mathbf p\) gives
\(\mathcal L^{(r,i)}[\mathbf F](\mathbf p)\preceq\mathbf F(\mathbf p)\).

\end{document}